\documentclass[aps,secnumarabic,amsmath,amssymb,nofootinbib,superscriptaddress,nobibnotes,twocolumn,showpacs]{revtex4}
\usepackage{float}
\usepackage[squaren]{SIunits}
\usepackage{verbatim}
\usepackage{mathrsfs}
\usepackage{comment}
\usepackage{braket}
\usepackage{bbold}
\usepackage[colorlinks]{hyperref}
\usepackage{pdfpages}
\usepackage{xcolor}
\usepackage[latin1]{inputenc}
\usepackage{graphicx}
\usepackage{amsmath}
\usepackage{dsfont}
\usepackage{amsfonts,amssymb}
\usepackage{amsthm}
\usepackage{xspace}
\usepackage{times}
\usepackage{longtable}
\usepackage{extarrows}
\usepackage{ulem} 
\hypersetup{colorlinks = true,linkcolor = violet,citecolor= blue}
\begin{document}

\title{Quantum simulation of gravitational-like waves in minisuperspace with an artificial qubit}
\date{\today}

\author{Ze-Lin Zhang}
\affiliation{Department of Physics, Fuzhou University, Fuzhou 350002, P. R. China}
\affiliation{Fujian Key Laboratory of Quantum Information and Quantum Optics, Fuzhou University, Fuzhou 350002, P. R. China}
\author{Ming-Feng Chen}
\affiliation{Department of Physics, Fuzhou University, Fuzhou 350002, P. R. China}
\affiliation{Fujian Key Laboratory of Quantum Information and Quantum Optics, Fuzhou University, Fuzhou 350002, P. R. China}
\author{Huai-Zhi Wu}
\affiliation{Department of Physics, Fuzhou University, Fuzhou 350002, P. R. China}
\affiliation{Fujian Key Laboratory of Quantum Information and Quantum Optics, Fuzhou University, Fuzhou 350002, P. R. China}
\author{Zhen-Biao Yang}
\email{The corresponding author's email address: zbyang@fzu.edu.cn}
\affiliation{Department of Physics, Fuzhou University, Fuzhou 350002, P. R. China}
\affiliation{Fujian Key Laboratory of Quantum Information and Quantum Optics, Fuzhou University, Fuzhou 350002, P. R. China}

\begin{abstract}
  On the background of the Born-Oppenheimer $($adiabatic$)$ approximation, we investigate the geometrical and topological structure in the theory of quantum gravity by using the path integral method and half-classical approximation. As we know, Berry curvature can be extracted from the linear response of a driven two-level system to nonadiabatic manipulations of its Hamiltonian. In parameter space of the Hamiltonian, magnetic monopoles can be artificially simulated. Ripples occur in Hilbert space when the monopole travels from inside to outside the surface of energy manifold spanned by system parameters. From this point of view, we set up the connection between the ripples characterized by the fidelity of quantum states in Hilbert space and gravitational-like waves in minisuperspace. This might open a window for the study of geometrical and topological properties of quantum gravity with the help of quantum physical systems.
\end{abstract}


\maketitle
\section{Introduction}
\label{sec: Introduction}

  It is recognized that quantum theory  and general relativity are two pillars of modern physics. However, the goal of finding a consistent theory of quantum gravity still remains elusive~\cite{Hamber-2009}. In spite of this, there still exist many ways to stimulate and guide researchers to solve problems of quantum cosmology, and one could find out the vulnerable spot of the difficulties in quantum cosmology through some already known characters of quantum gravity~\cite{Zee-1979,Teitelboim,Hartle-1983,Vilenkin-1985,Horowitz-1985,Maeda-1987,Liu-1988,Gott-1998,Ashtekar-2009}.

  Quantum gravity in path integral form was significantly stimulated by the superspace approach to the canonical quantization of gravity~\cite{Teitelboim,Hartle-1983}. A profound formulation of quantum mechanics of quantization of a system is based on the fact that the Feynman propagator can be written as a sum over all possible paths between the initial and final points in space-time~\cite{Feynman-1965}. In quantum gravity, one should sum over all possible paths linking the two given three-geometries and integrate over all the possible corresponding separations of given local proper times.

  As is known to us, time does not appear in the Hamiltonian form of gravity. This is a consequence of the invariance of the action $\textit{S}$ under arbitrary space-time transformations, from which follows $\delta \textit{S}/\delta \textit{g}_{\mu\nu} = 0$, where $\textit{g}_{\mu\nu}$ is the metric of space-time. We expect that such a feature will be maintained in a quantum formulation. From this point of view, R. Balbinot \textit{et al.}~\cite{Balbinot} proposed a system composed by matter and gravity in which the former follows the latter adiabatically in space-time of spatial lattice points. It is even more noteworthy that they used the method of the Born-Oppenheimer $($BO$)$ approximation~\cite{DCAJ-2004}. Actually, this idea was first introduced by T. Banks during the research of the semiclassical approximation to the quantum gravity wave function~\cite{Banks-1985}. It has been observed that the introduction of matter allows one to introduce the concept of time, and time parametrizes how matter follows gravity.  After that, R. Brout generalized Banks's procedure by grafting a BO-type approximation onto the semiclassical approximation of the gravitational part of the solution of the Wheeler-DeWitt equation~\cite{Brout-1987}. In particular, it has been noticed that the semiclassical wave function for gravity provides a parametrization for the evolution of matter in which the latter follows the former adiabatically. The ideas of all of these are that the mass scale of matter is $\ll$ $\textit{m}_{\textit{P}}$ $($the Planck mass$)$.

  A wave function acquires a geometric phase in addition to the dynamical one if the system Hamiltonian is modulated by a set of slowly varying external parameters~\cite{Berry-1984}. Towards the matter-gravity $($MG$)$ system, we may consider the coordinates associated with the former as being the fast $($light$)$ variables and those associated with the latter as being the slow $($heavy$)$ variables.  The excursion along a closed loop $\textit{C}$ in the external parameter space leads to an adiabatic phase that is a matter wave function of the slowly changing heavy degrees of freedom. During the process one passes near a point in which the matter state turns into double degeneracy $($for a two-level system$)$. With regard to the three-dimensional case, a monopolelike singularity can be used to study the properties of topological structure of the parameter space with the aid of the distribution of Berry curvature. In this paper, we acquire the Berry curvature by two different ways, the path integral method and half-classical approximation, all based on the BO approximation. It is noticed that when the light system, coupled to the heavy classical one, is quantal, this situation is what we call  ``half-classical mechanics'' $($there being no implication that the light system is semiclassical, i.e., near its classical limit$)$~\cite{MVB1993}. Based on the concept of quasiparticles appearing in the field of condensed matter physics, we propose a new concept of quasilattice points in space-time under the condition of half-classical approximation, and it is proved equivalent to the case of that in the method of path integral in calculating gravitational Berry curvature. The curvature can be extracted from the linear response of a driven two-level system to nonadiabatic manipulations of its Hamiltonian~\cite{VGAP-2012,SKKSGVPPL-2014,PR-2014}.

  Recently we have proposed a method of quantum simulation of the magnetic monopoles by using a driven superconducting qubit~\cite{Zhang-2016}, and shown that when the monopoles pass from inside to outside the Hamiltonian manifold, the quantum states generate ripples in the Hilbert space under the influence of Berry curvatures. In this work, we reconsider the general magnetic force produced by the monopole centered at the origin of coordinates spanned by a set of parameters in the Hamiltonian. In combination with all points of a spatial lattice $($all three-space$)$ and spatial metric tensor in space-time, we compare the singularity in minisuperspace $($which also can be viewed as a parameter space$)$ with the monopole in parameter space.

  It is well known that Einstein's theory of general relativity predicts the existence of gravitational waves that are ripples in space-time~\cite{Schutz-2009}, created in certain gravitational interactions that travel outward from their sources~\cite{BPA-2016}. The gravitational interactions, like the process of a binary black hole merger, could be viewed as the process of two singularities turning into one. The amplitude of gravitational waves in the merging instant reaches a maximum. In our heuristic analogy, this instant corresponds to the moment that ripples emerge in Hilbert space when the monopole travels through $($collision$)$ the energy surface of the Hamiltonian manifold.  As we know, the different number of singularities in physical systems represents the different topological structures~\cite{MN-1998}. This thus might open a window for the study of the geometrical and topological properties in quantum gravity with the help of quantum physical systems.

  In Sec.~\ref{sec: Effective action}, we introduce a path integral description of the interaction MG system in the adiabatic approximation. As a gauge connection associated with the Berry curvature is induced during a closed loop $\textit{C}$, we can obtain the nontrivial topological structure in the manifold of all points of the spatial lattice $($all three-metric$)$. In Sec.~\ref{sec: Half-classical}, We propose a notion of quasilattice points in space-time in the study of half-classical approximation, which is equivalent to the case of that in the path integral. This method provides the possibility to measure the Berry curvature in minisuperspace. As an analogy, in Sec.~\ref{sec: Ripples}, we utilize an effective method to directly measure the Berry curvature by way of a nonadiabatic response on physical observables to an external parameter's rate of change. Then we set up the connection between the ripples characterized by the fidelity of quantum states in Hilbert space and gravitational-like waves in minisuperspace. Finally, in Sec.~\ref{sec: Conclusion}, we discuss and summarize our results.

  \section{Effective path integral and quantum gravitational geometric tensor}
  \label{sec: Effective action}

  Now we first decompose the manifold of space-time into $\Sigma^{\textrm{3}}\times\mathbb{R}^{\textrm{1}}$ where $\Sigma^{\textrm{3}}$ denotes the three-dimensional manifold $($spacelike hypersurface $\textit{x}^{\textit{i}}$, $\textit{i}=\textrm{1,2,3}$$)$ and $\mathbb{R}^{\textrm{1}}$ denotes the real line $($time $\textit{t} = \textit{x}^{\textrm{0}}$$)$, and it cannot change upon evolution. The four-dimensional line element under the standard $\textrm{3+1}$ decomposition is~\cite{Hartle-1983,Hamber-2009}
  \begin{eqnarray}\label{4Dlineelement}
  {\textit{ds}}^{\textrm{2}} = -(\textit{N}^{\textrm{2}}-\textit{N}_{\textit{i}}\textit{N}^{\textit{i}})\textit{dt}^{\textrm{2}}+2\textit{N}_{\textit{i}}\textit{dx}^{\textit{i}}\textit{dt}+{\textit{h}}_{\textit{ij}}\textit{dx}^{\textit{i}}
  \textit{dx}^{\textit{j}},
  \end{eqnarray}
  where we use the induced metric ${\textit{h}}_{\mu\nu} = \textit{g}_{\mu\nu} + \textit{n}_{\mu}\textit{n}_{\nu}$; more details are shown in Appendix~\ref{sec: Proof1}. The total Hamiltonian is given by
  \begin{eqnarray}\label{Hamiltonian1}
  {\mathcal{H}} &=& {\mathcal{H}}^{\mathds{G}} + {\mathcal{H}}^{\mathds{M}} = \int\textit{d}^{\textrm{3}}\textit{x}(\textit{N}\mathcal{H}_{\textrm{0}}^{\mathds{F}}+\textit{N}^{\textit{i}}\mathcal{H}_{\textit{i}}^{\mathds{F}})\cr\cr
  &=& \sum_{\textit{x}} \lambda^{\textrm{3}}\big[(\textit{N}{\mathcal{H}_{\textrm{0}}^{\mathds{G}}}+ \textit{N}^{\textit{i}}{\mathcal{H}_{\textit{i}}^{\mathds{G}}}) + (\textit{N}{\mathcal{H}_{\textrm{0}}^{\mathds{M}}}+ \textit{N}^{\textit{i}}{\mathcal{H}_{\textit{i}}^{\mathds{M}}}) \big]\cr\cr
  &\equiv&\sum_{\textit{x}}{\mathcal{H}}_{\textit{x}},
  \end{eqnarray}
   where ${\mathcal{H}_{\textrm{0}}^{\mathds{G}}}$ and ${\mathcal{H}_{\textit{i}}^{\mathds{G}}}$ are the Hamiltonians associated with the gravitational field. Moreover, ${\mathcal{H}_{\textrm{0}}^{\mathds{M}}}$ and ${\mathcal{H}_{\textit{i}}^{\mathds{M}}}$ are stress-energy tensors of the matter field projected in the normal direction to the three-dimensional spacelike surface $[$with one component normal $($ $\textit{N}\mathcal{H}_{\textrm{0}}^{\mathds{M}}$$)$ and another tangential $($ $\textit{N}^{\textit{i}}\mathcal{H}_{\textit{i}}^{\mathds{M}}$$)$$]$. The integral over three-space is replaced by a sum over all points $\textit{x}$ of a spatial lattice of volume $\lambda^{\textrm{3}}$, thus obtaining a sum of Hamiltonians at every single lattice point. The lapse function $\textit{N}$ and shift vector $\textit{N}^{\textit{i}}$ play the role of Lagrange multipliers and satisfy the proper time gauge conditions $\partial \textit{N}/\partial \tau = \partial \textit{N}^{\textit{i}}/\partial \tau = \textrm{0}$.

  \begin{figure}[h]
    \centering
    \includegraphics[width=3.4in]{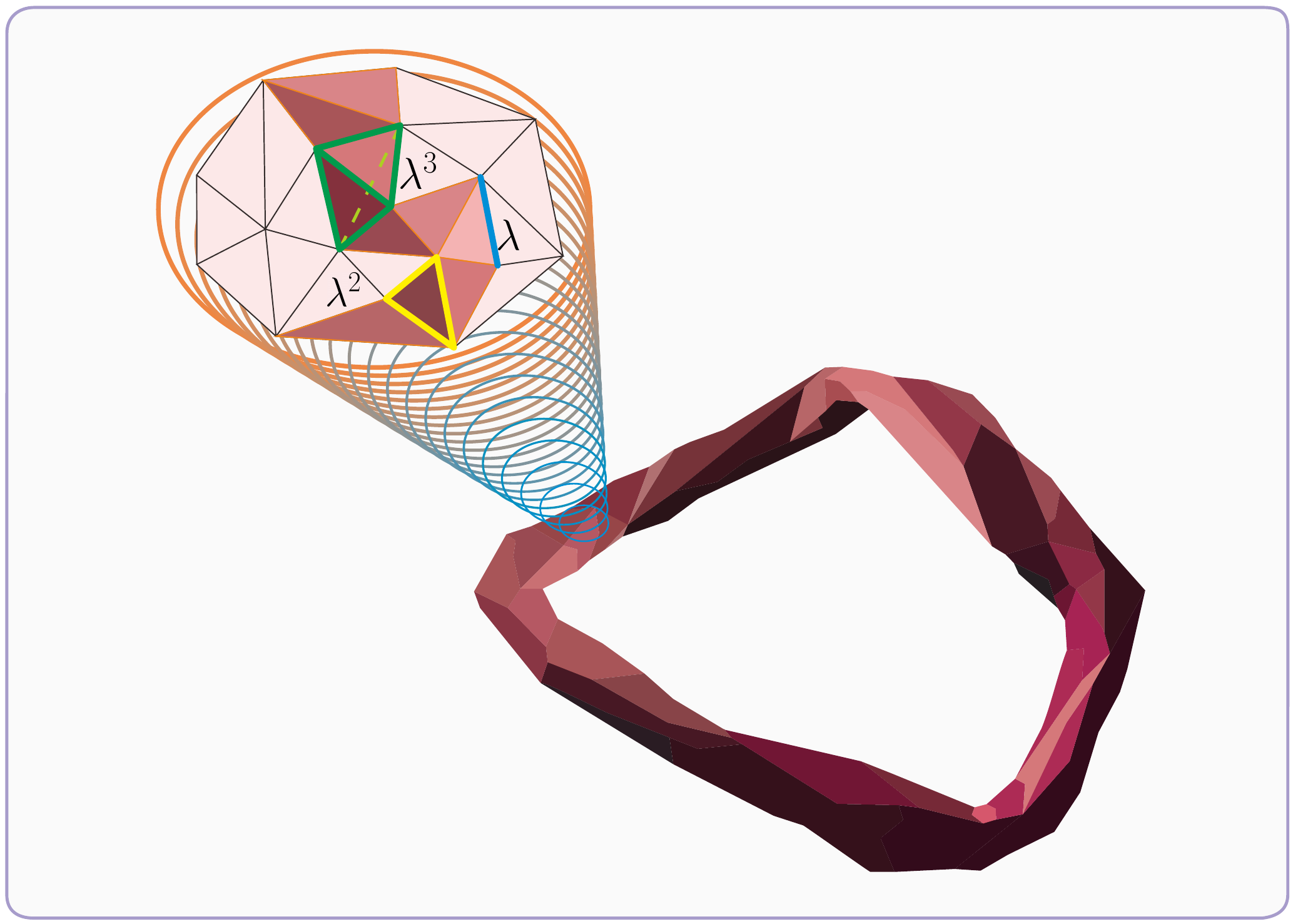}
    \caption{$($Color online$)$ Illustration of a spatial lattice. The elementary building blocks for three-dimensional space-time are simplices of dimension three. A one-simplex is an edge $($${\lambda}$, the blue line$)$, a two-simplex is a triangle $($${\lambda}^{\textrm{2}}$, the yellow lines$)$, and a three-simplex is a tetrahedron $($${\lambda}^{\textrm{3}}$, the green lines$)$.  On a random simplicial lattice there are in general no preferred directions, except for the special path (the maroon part) we choose here.}
    \label{fig:Figure_1}
  \end{figure}

  To get the wanted gravitational wave function, let us consider the new Hamiltonian in Hilbert space,
  \begin{eqnarray}\label{Hamiltonian2}
  \hat{{\mathcal{H}}} = \sum_{\textit{x}}\lambda^{\textrm{3}}\big[\textit{N}\hat{\mathcal{H}}_{\textrm{0}}^{\mathds{R}} + \textit{N}^{\textit{i}}({\hat{\mathcal{H}}_{\textit{i}}^{\mathds{G}}}+{\hat{\mathcal{H}}_{\textit{i}}^{\mathds{M}}})\big]\equiv\sum_{\textit{x}}\hat{{\mathcal{H}}}_{\textit{x}},
  \end{eqnarray}
  where $\hat{\mathcal{H}}_{\textrm{0}}^{\mathds{R}}={\hat{\mathcal{H}}_{\textrm{0}}^{\mathds{G}}}+{\hat{\mathcal{H}}_{\textrm{0}}^{\mathds{M}}} + {{\hbar}^{\textrm{2}}\mathcal{R}}/{\textrm{12}}$, and $\mathcal{R}$ is the curvature scalar related to the manifold of all three-metrics and has been introduced for the sake of taking into account operator ordering~\cite{DeWitt}.

 The MG state vectors of product form $|\phi({\textit{h}},\textit{x}),{\textit{h}}_{\textit{ij}}(\textit{x})\rangle\equiv|\phi({\textit{h}},\textit{x})\rangle\otimes
  |{\textit{h}}_{\textit{ij}}(\textit{x})\rangle$ where ${\textit{h}}_{\textit{ij}}(\textit{x})$ and $\phi({\textit{h}},\textit{x})$ separately describe the gravitational field and the matter field. The transition matrix element between the initial state $|\phi^{\circ}, {\textit{h}}^{\circ}\rangle$ and the final state $|\phi^{\bullet}, {\textit{h}}^{\bullet}\rangle$ can be written as
\begin{widetext}
  \begin{eqnarray}\label{transition matrix element}
  \langle\phi^{\bullet}, {\textit{h}}^{\bullet}|\phi^{\circ}, {\textit{h}}^{\circ}\rangle &=& \int\langle\phi^{\bullet}, {\textit{h}}^{\bullet}, \textit{N}^{\bullet}, \textit{N}^{{\bullet}\textit{i}}, {\textit{t}}^{\bullet}|\phi^{\circ}, {\textit{h}}^{\circ},  \textit{N}^{\circ}, \textit{N}^{{\circ}\textit{i}}, {\textit{t}}^{\circ}\rangle\prod_{\textit{x},\tau,\textit{i}}
  \textit{d}\big[\textit{N}({\textit{t}}^{\bullet}-{\textit{t}}^{\circ})\big]\textit{d}\big[\textit{N}^{\textit{i}}({\textit{t}}^{\bullet}-{\textit{t}}^{\circ})\big]\delta\big(\partial{\textit{N}}/\partial{\tau}
  \big)\delta\big(\partial{\textit{N}^{\textit{i}}}/\partial{\tau}\big)\cr
  &=& \int\langle\phi^{\bullet}, {\textit{h}}^{\bullet}|\big(\textrm{exp}\big[-\textit{i}\epsilon\hat{{\mathcal{H}}}/\hbar\big]\big)^{\textit{N}}|\phi^{\circ}, {\textit{h}}^{\circ}\rangle\prod_{\textit{x},\textit{i}}\textit{d}{\textit{t}^{\textit{i}}}\textit{d}{\textit{t}},
  \end{eqnarray}
  where $\epsilon={\textit{T}}/{\textit{N}}$ $($\textit{T} denotes the whole evolution time and \textit{N} is a large number$)$, and ${\hat{\mathcal{H}}}$ is the Hamiltonian of the MG system in Hilbert space; also the invariant interval between two adjacent lattice points equals $\textit{N}^{\bullet}{\textit{t}}^{\bullet}-\textit{N}^{\circ}{\textit{t}}^{\circ}$ $($$\textit{N}^{{\bullet}\textit{i}}{\textit{t}}^{\bullet}-\textit{N}^{{\circ}\textit{i}}{\textit{t}}^{\circ}$$)$. As shown in Fig.~\ref{fig:Figure_1}, we have divided the interval into \textit{N} equal segments of size $\lambda({\lambda}^{\textit{i}})$ so that $\lambda = (\textit{N}^{\bullet}{\textit{t}}^{\bullet}-\textit{N}^{\circ}{\textit{t}}^{\circ})/\textit{N}$. The proper time gauge conditions allow us to introduce a local proper time $\textit{t}=\textit{N}({\textit{t}}^{\bullet}-{\textit{t}}^{\circ})$ with $\textit{N}=\textit{N}^{\circ}=\textit{N}^{\bullet}$ at every single spatial point $($similarly for ${\textit{t}}^{\textit{i}}$$)$,
  \begin{eqnarray}\label{transition matrix element1}
  \langle\phi^{\bullet}, {\textit{h}}^{\bullet}|\phi^{\circ}, {\textit{h}}^{\circ}\rangle = \int\langle\phi^{\bullet}, {\textit{h}}^{\bullet}|\textrm{exp}[-\textit{i}\epsilon\hat{\mathcal{H}}(\textit{N})/\hbar]|\textit{p}(\textit{N})\rangle\langle\textit{p}(\textit{N})|\textit{q}(\textit{N}-\textrm{1})\rangle\langle\textit{q}
  (\textit{N}-\textrm{1})|\textrm{exp}[-\textit{i}\epsilon\hat{\mathcal{H}}(\textit{N}-\textrm{1})/\hbar] \cr
  \times\cdots|\textit{q}(\textrm{1})\rangle\langle\textit{q}(\textrm{1})|\textrm{exp}[-\textit{i}\epsilon\hat{\mathcal{H}}(\textrm{1})/\hbar]|\textit{p}(\textrm{1})\rangle\langle\textit{p}(\textrm{1})|\phi^{\circ}, {\textit{h}}^{\circ}\rangle\prod_{\textit{x},\textit{i},\textit{L}}\textit{d}{\textit{t}^{\textit{i}}}\textit{d}{\textit{t}}~\frac{\textit{d}{\textit{p}}_{\textit{L}}(\textit{N})}{\textrm{2}\pi\hbar}
  \prod_{\textit{k}=\textrm{1}}^{\textit{N}-\textrm{1}}\textit{d}{{\textit{q}}^{\textit{L}}}(\textit{k})\frac{\textit{d}{\textit{p}}_{\textit{L}}(\textit{k})}{\textrm{2}\pi\hbar},
  \end{eqnarray}
 where $\hat{\mathcal{H}}(\textit{k})$ describes the Hamiltonian at a lattice point $\textit{k}\lambda$ on the evolution path excursing from ${\textit{h}}^{\circ}$ to ${\textit{h}}^{\bullet}$. We used the completeness relation for ${\textit{q}}^{\textit{L}}$ and its conjugate momentum ${\textit{p}}_{\textit{L}}=-\textit{i}\hbar(\delta/\delta{{\textit{q}}^{\textit{L}}})$ here~\cite{Capital}.
 When $\textit{N}$ tends to infinity, the transition matrix element then becomes
 \begin{eqnarray}\label{transition matrix element2}
 \langle\phi^{\bullet}, {\textit{h}}^{\bullet}|\phi^{\circ}, {\textit{h}}^{\circ}\rangle = \int\textrm{exp}[\textit{i}{\textit{S}}(\textit{h})/\hbar]\textit{G}(\phi^{\bullet},\phi^{\circ})
 \prod_{\textit{L}}\frac{\textrm{1}}{\textrm{2}\pi\hbar}\mathcal{D}[{\textit{p}}_{\textit{L}}]\mathcal{D}[{\textit{q}}^{\textit{L}}]\prod_{\textit{x},\textit{i}}\textit{d}{\textit{t}^{i}}\textit{d}{\textit{t}},
 \end{eqnarray}
 where the path integral measure can be denoted as
 \begin{eqnarray}\label{integral measure}
 \int\frac{\mathcal{D}[{\textit{p}}_{\textit{L}}]}{\textrm{2}\pi\hbar}\mathcal{D}[{\textit{q}}^{\textit{L}}] = \lim_{\textit{n}\to\infty}\prod_{\textit{k}=\textrm{1}}^{\textit{N}-\textrm{1}}
 \bigg[\int\textit{d}{{\textit{q}}^{\textit{L}}}(\textit{k})\bigg]\prod_{\textit{k}=\textrm{1}}^{N}\bigg[\int\frac{\textit{d}{\textit{p}}_{\textit{L}}(\textit{k})}
 {\textrm{2}\pi\hbar}\bigg],
 \end{eqnarray}
 and the amplitude for the transition as the matter at a lattice point $\textit{k}\lambda$ excursing along the path from ${\textit{h}}^{\circ}$ to ${\textit{h}}^{\bullet}$ can be written as
 ~\cite{PP-1969,HKSI-1985}
 \begin{eqnarray}\label{amplitude}
 \textit{G}(\phi^{\bullet},\phi^{\circ}) = \langle\phi^{\bullet}|\textrm{exp}[-\textit{i}\epsilon\hat{\mathcal{H}}^{\mathds{M}}(\textit{N})/\hbar]\cdots\textrm{exp}[-\textit{i}\epsilon \hat{\mathcal{H}}^{\mathds{M}}(\textit{k})/\hbar]\cdots\textrm{exp}[-\textit{i}\epsilon \hat{\mathcal{H}}^{\mathds{M}}(\textrm{1})/\hbar]|\phi^{\circ}\rangle,
 \end{eqnarray}
 where $\hat{\mathcal{H}}^{\mathds{M}}(\textit{N})$ is the Hamiltonian of matter, and the gravitational action ${\textit{S}}(\textit{h})$ reads
 \begin{eqnarray}\label{gravitational action}
 {\textit{S}}(\textit{h}) = \int\textit{d}^{\textrm{3}}{\textit{x}}\bigg[{\int}_{{\textit{h}}^{\circ}}^{{\textit{h}}^{\bullet}}{{\textit{p}}_{\textit{L}}}\delta{{\textit{q}}^{\textit{L}}}-\bigg(
 {\int}_{{\textit{t}}^{\circ}}^{{\textit{t}}^{\bullet}}\delta{\textit{t}}~\hat{\mathcal{H}}_{\textrm{0}}^{\mathds{R}}+{\int}_{{\textit{t}}^{\circ\prime}}^{{\textit{t}}^{\bullet
 \prime}}\delta{\textit{t}^{\prime}}~\hat{\mathcal{H}}_{\textit{i}}^{\mathds{G}}\bigg)\bigg].
 \end{eqnarray}

 The matter states at every single lattice point $\textit{k}$ on the path could be described as follows:
 \begin{eqnarray}\label{amplitude2}
 \textit{G}(\phi^{\bullet},\phi^{\circ}) = \sum_{\phi(\textit{k}=\textrm{1})}^{\phi(\textit{N}-\textrm{1})}\langle\phi^{\bullet}|\textrm{exp}[-\textit{i}\epsilon\hat{\mathcal{H}}^{\mathds{M}}(\textit{N})/\hbar]|\phi(\textit{N}-\textrm{1})\rangle\langle
 \phi(\textit{N}-\textrm{1})|\cdots|\phi(\textrm{1})\rangle\langle\phi(\textrm{1})|\textrm{exp}[-\textit{i}\epsilon \hat{\mathcal{H}}^{\mathds{M}}(\textrm{1})/\hbar]|\phi^{\circ}\rangle,
 \end{eqnarray}
\end{widetext}
 where we used the completeness relation of matter states $\sum_{\textit{n}=\textrm{1}}^{\textit{N}}|\phi(\textit{n})\rangle\langle\phi(\textit{n})|=\textit{I}$.

 Since the mass scale related to gravity $($Planck mass, $\thicksim \textrm{10}^{-\textrm{6}}\textit{g}$$)$ is much larger than that related to usual matter, we could suppose that the latter follows the former adiabatically. In the BO approximation, the matter states might be considered to perform the adiabatic motion in the same quantum number $\phi=\phi^{\circ}=\phi^{\bullet}$~\cite{Balbinot}. Now we introduce the notion of energy density related to an adiabatic level $\phi$ at ${\textit{q}}^{\textit{L}}={\textit{q}}^{\textit{L}}(\textit{k})$. The Schr\"{o}dinger equation of matter states could be written as
 \begin{eqnarray}\label{Hamiltionian3}
 \hat{\mathcal{H}}_{\textrm{0}}^{\mathds{M}}(\textit{k})|\phi(\textit{k})\rangle = \mathcal{E}^{\phi}(\textit{k})|\phi(\textit{k})\rangle.
 \end{eqnarray}
 Then we have
\begin{widetext}
 \begin{eqnarray}\label{amplitude3}
 \textit{G}(\phi,\phi) = \textrm{exp}\big[-\frac{\textit{i}}{\hbar}\sum_{\textit{x}}\sum_{\textit{k}}\lambda^{\textrm{3}}\big(\lambda\mathcal{E}^{\phi}(\textit{k})+\lambda^{\textit{i}}\langle\phi|
 \hat{\mathcal{H}}_{\textit{i}}^{\mathds{M}}|\phi\rangle\big)\big]\lim_{\textit{N}\to\infty}\prod_{\textit{k}=\textrm{1}}^{\textit{N}}\langle\phi(\textit{k})|\phi(\textit{k}-\textrm{1})\rangle.
 \end{eqnarray}
\end{widetext}
 We notice that each factor $\langle\phi(\textit{k})|\phi(\textit{k}-\textrm{1})\rangle$ in Eq.~$($\ref{amplitude3}$)$ defines a connection between two infinitesimally separated points $\phi(\textit{k}-\textrm{1})$ and $\phi(\textit{k})$. Therefore, Eq.~$($\ref{amplitude3}$)$ gives a finite connection along a path given by a set of discrete points $\{\textit{k}\}$. Thus, by using the Taylor expansion and expanding to the first order, we obtain
 \begin{eqnarray}\label{connection}
 \langle\phi(\textit{k})|\phi(\textit{k}-\textrm{1})\rangle &\simeq& \textrm{1}-\sum_{\textit{x}}\lambda^{\textrm{3}}\langle\phi(\textit{k})|\delta/\delta{{\textit{q}}^{\textit{L}}}|
 \phi(\textit{k})\rangle\delta{{\textit{q}}^{\textit{L}}}\cr\cr
 &\approx& \textrm{exp}\big[\textit{i}\sum_{\textit{x}}\lambda^{\textrm{3}}\langle\phi(\textit{k})\big|\textit{i}\delta/\delta{{\textit{q}}^{\textit{L}}}\big|\phi(\textit{k})\rangle\delta{{\textit{q}}^{\textit{L}}}\big] \cr\cr
 &=& \textrm{exp}[\textit{i}\omega/\hbar].
 \end{eqnarray}
 From Eq.~$($\ref{amplitude3}$)$, we suppose that
 \begin{eqnarray}\label{geometric phase1}
 \lim_{\textit{N}\to\infty}\prod_{\textit{k}=\textrm{1}}^{\textit{N}}\langle\phi(\textit{k})|\phi(\textit{k}-\textrm{1})\rangle
 &=& \langle\phi({\textit{t}}^{\bullet})|\phi({\textit{t}}^{\circ})\rangle \cr\cr
 &=& \textrm{exp}[{\textit{i}\Omega_{\phi}}/{\hbar}]
 \end{eqnarray}
 with
 \begin{eqnarray}\label{geometric phase2}
 \Omega_{\phi} = \int_{\textit{\textit{h}}^{\circ}}^{\textit{\textit{h}}^{\bullet}}\omega = \int_{\textit{\textit{h}}^{\circ}}^{\textit{\textit{h}}^{\bullet}}
 \delta{{\textit{q}}^{\textit{L}}}\langle\phi|\textit{i}\hbar\delta/\delta{{\textit{q}}^{\textit{L}}}|\phi\rangle.
 \end{eqnarray}

 For simplicity, still in the adiabatic approximation, we let the wave function of matter evolve along a closed loop $\textit{C}$ in superspace, and $\phi(\textrm{0})=\phi({\textit{\textit{t}}^{\circ}})$, $ \phi(\textit{T})=\phi({\textit{\textit{t}}^{\bullet}})$; then Eq.~$($\ref{geometric phase2}$)$ turns into
 \begin{eqnarray}\label{geometric phase22}
 \Omega_{\phi}(\textit{C}) = \oint_{\textit{C}}\omega = \oint_{\textit{C}}\delta{{\textit{q}}^{\textit{L}}}\langle\phi|\textit{i}\hbar\delta/\delta{{\textit{q}}^{\textit{L}}}|\phi\rangle.
 \end{eqnarray}
 This form is essentially the same as the Berry phase $($actually, this is a variational version of the Berry phase$)$. Therefore, we obtain the effective transition amplitude related to the adiabatic change of the external dynamical variable $\phi$,
\begin{widetext}
 \begin{eqnarray}\label{amplitude4}
 {\textit{G}}_{\textrm{eff}}(\textit{T}) = \textrm{exp}\bigg[-\frac{\textit{i}}{\hbar}\int\textit{d}^{\textrm{3}}\textit{x}\bigg(\int_{\textrm{0}}^{\textit{T}}\delta{\textit{t}}\langle\phi|\hat{\mathcal{H}}_{\textrm{0}}^{\mathds{M}}|\phi\rangle
 +\int_{\textrm{0}}^{\textit{T}^{\prime}}\delta{\textit{t}}^{\prime}\langle\phi|\hat{\mathcal{H}}_{\textit{i}}^{\mathds{M}}|\phi\rangle-
 \oint_{\textit{C}}\delta{{\textit{q}}^{\textit{L}}}\langle\phi|\textit{i}\hbar\delta/\delta{{\textit{q}}^{\textit{L}}}|\phi\rangle\bigg)\bigg].
 \end{eqnarray}
 The effective path integral for the transition matrix element is given by
 \begin{eqnarray}\label{amplitude5}
 \langle\phi^{\bullet}, {\textit{h}}^{\bullet}|\phi^{\circ}, {\textit{h}}^{\circ}\rangle_{\textrm{eff}} = \int\textrm{exp}\bigg[\frac{\textit{i}}{\hbar}\bigg({\textit{S}}_{\textit{ad}}+\int\textit{d}^{\textrm{3}}\textit{x}\int_{{\textit{h}}^{\circ}}^{{\textit{h}}^{\bullet}}
 \delta{{\textit{q}}^{\textit{L}}}\langle\phi|\textit{i}\hbar\delta/\delta{{\textit{q}}^{\textit{L}}}|\phi\rangle\bigg)\bigg]\prod_{\textit{L}}\frac{\textrm{1}}{\textrm{2}\pi\hbar}\mathcal{D}
 [{\textit{p}}_{\textit{L}}]\mathcal{D}[{\textit{q}}^{\textit{L}}]\prod_{\textit{x},\textit{i}}\textit{d}{\textit{t}^{\textit{i}}}
 \textit{d}{\textit{t}}
 \end{eqnarray}
 with the adiabatic action function~\cite{HKSI-1985}
 \begin{eqnarray}\label{adiabatic action}
 {\textit{S}}_{\textit{ad}} = {\textit{S}}(\textit{h})-\int\textit{d}^{\textrm{3}}\textit{x}\bigg(\int_{\textit{t}^{\circ}}^{\textit{t}^{\bullet}}\delta\textit{t}~{\mathcal{E}}^{\phi}+\int_{\textit{t}^{\circ\prime}}^{\textit{t}^{\bullet
 \prime}}\delta\textit{t}^{\textit{i}}\langle\phi|\hat{\mathcal{H}}_{\textit{i}}^{\mathds{M}}|\phi\rangle\bigg).~~~~~
 \end{eqnarray}
\end{widetext}
 The phase $\Omega_{\phi}(\textit{C})$ appears as a topological action function in the MG system. This implies that the eigenstate of matter that is only the function of the three-metric $\textit{q}^{\textit{L}}$ stays the same during the motion of the system in superspace. In the presence of an effective gauge field that can be descried by the ``vector potential'' $\mathcal{A}=\langle\phi|\textit{i}\hbar\delta/\delta{{\textit{q}}^{\textit{L}}}|\phi\rangle$, one might have nontrivial topological structure in the manifold of superspace in case one allows for a gauge transformation of it. The connection $\mathcal{A}$ is substantially the Berry-Simon connection~\cite{Simon-1983}.

 According to the adiabatic condition, the matter states always stay in the same quantum number in Eq.~$($\ref{amplitude3}$)$. Clearly the lowest correction to this is actually to allow just one intermediate transition to another eigenstate subsequently returning to the initial eigenstate. Thus the transition matrix element, also called the quantum gravitational geometric tensor, reads
 \begin{eqnarray}\label{lowest correction}
 \mathcal{G}_{\textit{XY}}^{\phi} &=& \sum_{\phi^{\prime}\neq\phi}\big\langle\phi\big|\frac{\delta}{\delta{\textit{q}^{\textit{X}}}}\big|\phi^{\prime}\big\rangle
 \big\langle\phi^{\prime}\big|\frac{\delta}{\delta{\textit{q}^{\textit{Y}}}}\big|\phi\big\rangle\cr\cr
 &=& -\sum_{\phi^{\prime}\neq\phi}\frac{\textrm{1}}{(\mathcal{E}^{\phi}-\mathcal{E}^{\phi^{\prime}})^{\textrm{2}}}\big\langle\phi\big|\frac{\delta\hat{\mathcal{H}}_{\textrm{0}}^{\mathds{M}}}
 {\delta{\textit{q}^{\textit{X}}}}\big| \phi^{\prime} \big\rangle\big\langle\phi^{\prime}\big|\frac{\delta\hat{\mathcal{H}}_{\textrm{0}}^{\mathds{M}}}{\delta{\textit{q}^{\textit{Y}}}}\big|\phi\big\rangle,\cr&&
 \end{eqnarray}
 where $\mathcal{G}_{\textit{XY}}^{\phi}$ is Hermitian and we note that the eigenfunction of the matter acquires an opposite adiabatic phase to the gravitational wave function. The transition matrix element in Eq.~$($\ref{lowest correction}$)$ could be decomposed into real symmetric and imaginary antisymmetric parts~\cite{JPGV-1980},
 \begin{eqnarray}\label{decomposed1}
 \textrm{Re}\mathcal{G}_{\textit{XY}}^{\phi}=\frac{\textrm{1}}{\textrm{2}}(\mathcal{G}_{\textit{XY}}^{\phi}+\mathcal{G}_{\textit{YX}}^{\phi}),~~
 \textrm{Im}\mathcal{G}_{\textit{XY}}^{\phi}=\frac{\textrm{1}}{\textrm{2}}(\mathcal{G}_{\textit{XY}}^{\phi}-\mathcal{G}_{\textit{YX}}^{\phi}).~~~~~~
 \end{eqnarray}
 The real part of the transition matrix element provides a possible method to measure the distance between two neighboring matter wave functions along paths in superspace $($parameter space$)$. The imaginary part in Eq.~$($\ref{decomposed1}$)$ can be related to the Berry-Simon connection. It can be written as
 \begin{eqnarray}\label{imaginary_part}
 \textrm{2}\textrm{Im}\mathcal{G}_{\textit{XY}}^{\phi}&=&\frac{\delta}{\delta{\textit{q}^{\textit{X}}}}\big\langle\phi\big|\frac{\delta}{\delta{\textit{q}^{\textit{Y}}}}\big|\phi
 \big\rangle-\frac{\delta}{\delta{\textit{q}^{\textit{Y}}}}\big\langle\phi\big|\frac{\delta}{\delta{\textit{q}^{\textit{X}}}}\big|\phi\big\rangle\cr\cr
 &=&-\sum_{\phi^{\prime}\neq\phi}\frac{\textrm{1}}{(\mathcal{E}^{\phi}-\mathcal{E}^{\phi^{\prime}})^{\textrm{2}}}
 \big[\big\langle\phi\big|\frac{\delta\hat{\mathcal{H}}_{\textrm{0}}^{\mathds{M}}}{\delta{\textit{q}^{\textit{X}}}}\big|\phi^{\prime} \big\rangle\big\langle\phi^{\prime}\big|\frac{\delta\hat{\mathcal{H}}_{\textrm{0}}^{\mathds{M}}}{\delta{\textit{q}^{\textit{Y}}}}\big|\phi\big\rangle\cr\cr
 &&-\big\langle\phi\big|\frac{\delta\hat{\mathcal{H}}_{\textrm{0}}^{\mathds{M}}}{\delta{\textit{q}^{\textit{Y}}}}\big|\phi^{\prime} \big\rangle\big\langle\phi^{\prime}\big|\frac{\delta\hat{\mathcal{H}}_{\textrm{0}}^{\mathds{M}}}{\delta{\textit{q}^{\textit{X}}}}\big|\phi\big\rangle\big],~~~~
 \end{eqnarray}
 which is a phase $\textrm{2}$-form in superspace, also called the gravitational Berry curvature. It indicates that degeneracies are some singularities $($micro black holes~\cite{bhmmm}$)$ that contribute nonzero terms to topological invariants~\cite{VGAP-2012,SKKSGVPPL-2014}. From Eqs.~$($\ref{amplitude3}$)$ to~$($\ref{amplitude4}$)$, the gravitational Berry phase can be given by
\begin{widetext}
 \begin{eqnarray}\label{BerryLattice}
 \int\textit{d}^{\textrm{3}}\textit{x}\oint_{\textit{C}}\delta{{\textit{q}}^{\textit{L}}}\langle\phi|\textit{i}\hbar\delta/\delta{{\textit{q}}^{\textit{L}}}|\phi\rangle
 =-\int\textit{d}^{\textrm{3}}\textit{x}\iint_{\sigma}\delta{\sigma}^{\textit{XY}}~\textrm{Im}\sum_{\phi^{\prime}\neq\phi}\frac{\big\langle\phi\big|{\delta\hat{\mathcal{H}}_{\textrm{0}}^{\mathds{M}}}/
 {\delta{\textit{q}^{\textit{X}}}}\big|\phi^{\prime} \big\rangle\big\langle\phi^{\prime}\big|{\delta\hat{\mathcal{H}}_{\textrm{0}}^{\mathds{M}}}/{\delta{\textit{q}^{\textit{Y}}}}\big|\phi\big\rangle}{(\mathcal{E}^{\phi}-\mathcal{E}^{\phi^{\prime}})^{\textrm{2}}},
 \end{eqnarray}
\end{widetext}
 where the summation in superspace is over the three-geometries on the spacelike three-surface and $\delta{\sigma}^{\textit{XY}}$ is the element of a two-sphere in superspace bounded by the loop $\textit{C}$.

  \begin{figure}[h]
    \centering
    \includegraphics[width=3.36in]{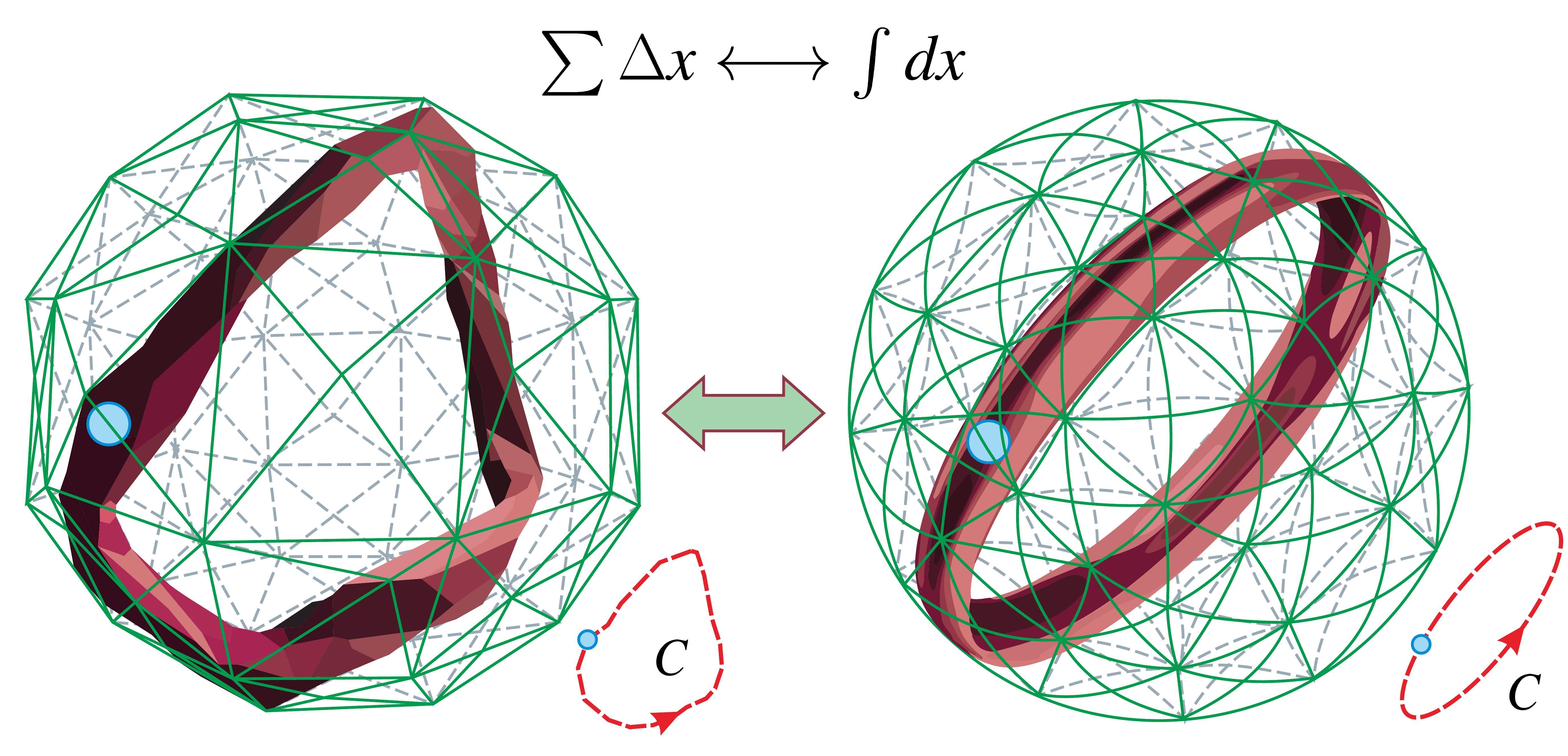}
    \caption{$($Color online$)$ Polyhedral approximation to a sphere. The green spatial lattice in the left is a closed three-dimensional manifold with all three-space $($sum over all points of a spatial lattice$)$, which is homeomorphic to a connected polyhedron in piecewise linear space, and its points possess neighborhoods that are homeomorphic to the interior of the three-dimensional sphere $($in the right$)$.  According to the Gauss-Bonnet theorem~\cite{PP-2006}, the topological structure between the polyhedra and the sphere is $($both have the same Euler characteristic$)$. The light blue ball represents the matter states associated with the closed red trajectories $\textit{C}$ of an adiabatic evolution. The dashed red line is the simplification of the red cyclic polyhedron $($all possible paths with the same quantum number corresponding to a net at each time, not a point$)$, which is homeomorphic to the red torus in the right.}
    \label{fig:Figure_2}
  \end{figure}

  Actually, Eq.~$($\ref{BerryLattice}$)$ could be depicted in a more comprehensible way in a form of Berry phase $\oint_{\textit{C}}\vec{\textit{A}}\cdot \textit{d}\vec{\textit{l}}= -\iint_{\textit{S}}\vec{\textit{B}}\cdot \textit{d}\vec{\textit{S}}$, where we used Stokes's theorem and $\nabla\times\vec{\textit{A}}=\vec{\textit{B}}$. The magnetic vector potential $\vec{\textit{A}}$ can be viewed as a vector form of the Berry-Simon connection we mentioned before; $\vec{\textit{B}}$ is a vector known as the Berry curvature analogous to the magnetic field in electromagnetism~\cite{Berry-1984}. Generally speaking, tensors are more general mathematical objects that can be used to represent the Berry curvature $\textit{F}_{\mu\nu}$; then Eq.~$($\ref{imaginary_part}$)$ becomes a more simple form,
  \begin{eqnarray}\label{Berry curvature1}
  \textit{F}_{\mu\nu} &=& \partial_{\mu}\textit{A}_{\nu}-\partial_{\nu}\textit{A}_{\mu}\cr\cr
   &=& -\textrm{Im}\sum_{n\neq m}\frac{\langle \psi_{m}|\partial_{\mu}\hat{H}|\psi_{n}\rangle\langle \psi_{n}|\partial_{\nu}\hat{H}|\psi_{m}\rangle -(\nu\leftrightarrow\mu)}{(\textit{E}_n-\textit{E}_m)^2},\cr&
  \end{eqnarray}
  where $\textit{A}_{\mu(\nu)}=\textit{i}\langle \psi_{m}|\partial_{\mu(\nu)}|\psi_{m}\rangle$ is just the Berry-Simon connection. However, in this context, the Berry phase is achieved by the method of path integral in the MG system on the background of the gravitational field $($not the electromagnetic field$)$, so we call it the gravitational Berry phase. As shown in Fig.~\ref{fig:Figure_2}, the gravitational Berry phase is under the background of a continuous spatial lattice form $($the transition from a smooth triangulation of a sphere to the corresponding secant approximation$)$. In Fig.~\ref{fig:Figure_2}, the evolution trajectory of the matter state on the spatial lattice sphere is a big circle $($geodesic line$)$.

\section{Half-classical adiabatic reactions of the matter-gravity system}
\label{sec: Half-classical}

  The method of adiabatic approximation has been studied in many different contexts. For example, in the Born-Oppenheimer studies of molecules, the dynamics of a composite physical system could be divided into two parts: the fast $($light$)$ system, which is described by fast variables $($e.g., the position $\textbf{\textit{q}}$ and momenta $\textbf{\textit{p}}$ of the electrons$)$; and slow $($heavy$)$ system, which is described by slow variables $($e.g., the position $\textbf{\textit{Q}}$ and momenta $\textbf{\textit{P}}$ of the nuclei$)$. The reasons for the BO approximation are readily revealed by studying gravity in the cosmological context.

  \begin{figure}[h]
    \centering
    \includegraphics[width=3.38in]{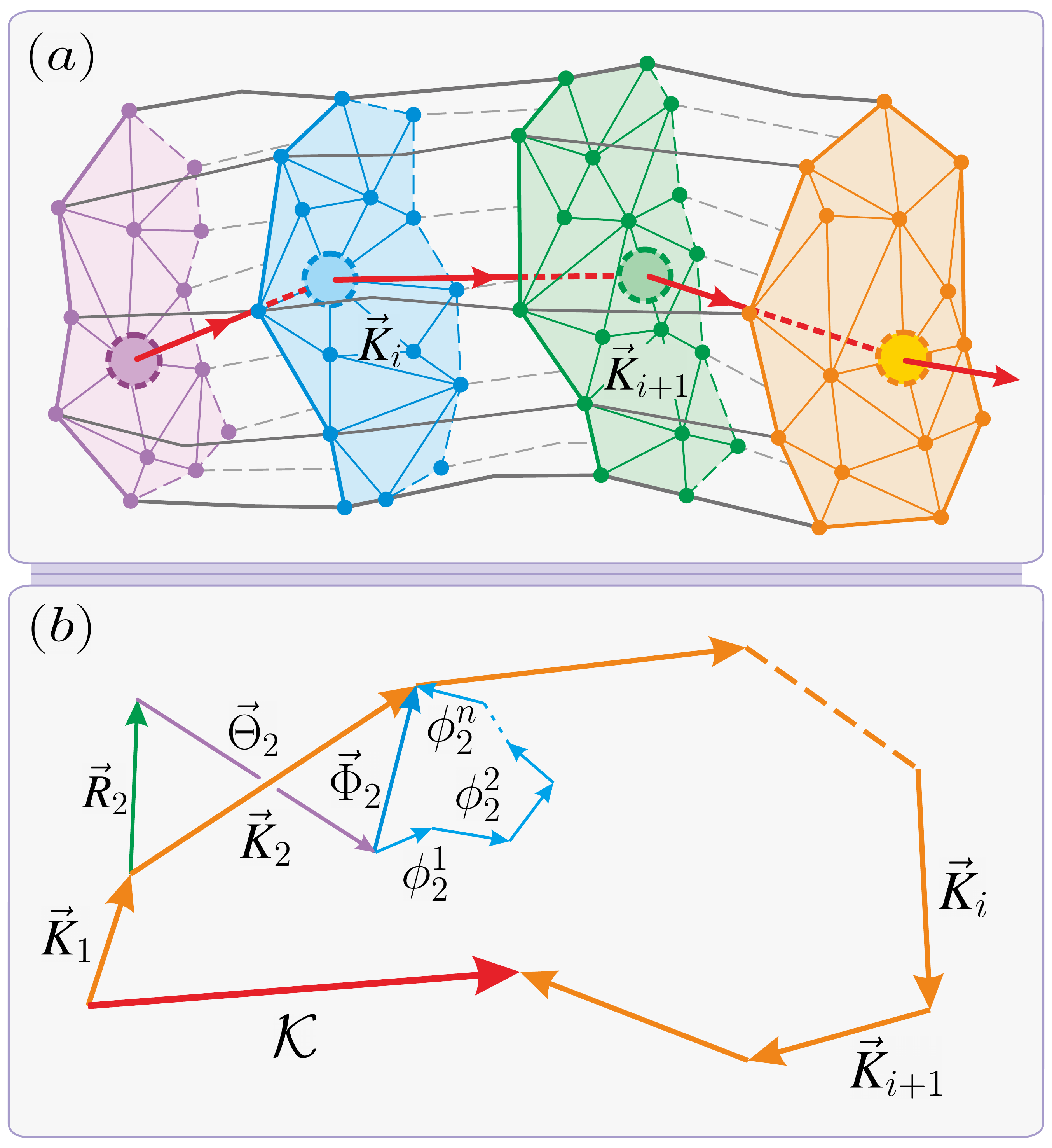}
    \caption{$($Color online$)$ Quasilattice points in space-time. $($a$)$ The trajectory of lattice points depicts the evolution history of the matter state in the gravitational field. The simultaneously created lattice points constitute a spatial net at every moment. Points on every spatial net can be viewed as equivalent to a quasilattice point on the trajectory of evolution $($dashed circles$)$. So the position of an ensemble of lattice points could be described by a set of quasilattice vectors. The quasilattice point from one to another reads $\vec{\textit{K}}_{\textit{i}}$ to $\vec{\textit{K}}_{\textit{i}+\textrm{1}}$. $($b$)$ Quasilattice function $\mathcal{K}$ in vector space. For example, we set $\vec{\textit{X}}_{\textit{n}}=\vec{\textit{R}}_{\textit{n}}, \vec{\textit{Y}}_{\textit{n}}=\vec{\Theta}_{\textit{n}}, \vec{\textit{Z}}_{\textit{n}}=\vec{\Phi}_{\textit{n}}$, and $\textit{x}_{\textit{n}}^{\textit{n}}=\textit{r}_{\textit{n}}^{\textit{n}}, \textit{y}_{\textit{n}}^{\textit{n}}=\theta_{\textit{n}}^{\textit{n}}, \textit{z}_{\textit{n}}^{\textit{n}}={\phi}_{\textit{n}}^{\textit{n}},$ $($here the subscript $\textit{n}={\textrm{2}}$$)$~\cite{string}.}
    \label{fig:Figure_3}
  \end{figure}

  In the matter-gravity system, we could consider the coordinates associated with the former as being the fast variables and those associated with the latter as being the slow variables. In order to describe the positions associated with the matter state, we first define the position of a single quasilattice point that could be considered as a set of quasilattice vectors of space-time. As shown in Fig.~\ref{fig:Figure_3}$($a$)$, the quasilattice function is defined as
  \begin{eqnarray}\label{position}
  \mathcal{K} &=& \ddagger\big\{\{\vec{\textit{K}}_{\textit{i}}\}\big\}\ddagger= \ddagger\big\{\{\vec{\textit{K}}_{\textrm{1}}\},..., \{\vec{\textit{K}}_{\textit{n}}\}\big\}\ddagger\cr\cr
  &=& \ddagger\big\{\{\vec{\textit{X}}_{\textrm{1}}, \vec{\textit{Y}}_{\textrm{1}}, \vec{\textit{Z}}_{\textrm{1}},...\},...,\{\vec{\textit{X}}_{\textit{n}}, \vec{\textit{Y}}_{\textit{n}}, \vec{\textit{Z}}_{\textit{n}},...\}\big\}\ddagger\cr\cr
  &=& \big\{\{\vec{\textit{X}}_{\textrm{1}},...,\vec{\textit{X}}_{\textit{n}}\}, \{\vec{\textit{Y}}_{\textrm{1}},...,\vec{\textit{Y}}_{\textit{n}}\}, \{\vec{\textit{Z}}_{\textrm{1}},...,\vec{\textit{Z}}_{\textit{n}}\},...\big\}\cr\cr
  &=& \big\{\{\vec{\textit{K}}^{\textrm{1}}\},..., \{\vec{\textit{K}}^{\textit{n}}\}\big\} = \big\{\{\vec{\textit{K}}^{\textit{i}}\}\big\},
  \end{eqnarray}
  where $\textit{i}=1,...,\textit{n}$, and the operator $\ddagger\big\{\{\}\big\}\ddagger$ is called the normal emergent set and can be transformed into the corresponding emergent set $\big\{\{\}\big\}$. $\vec{\textit{K}}_{\textrm{1}}$ and $\vec{\textit{X}}_{\textrm{1}}$ belong to different levels of space, respectively~\cite{LAA-2011}. The function $\mathcal{K}$ can be viewed as a set of quasilattice vectors $\big\{\{\vec{\textit{K}}_{\textrm{1}}\},..., \{\vec{\textit{K}}_{\textit{n}}\}\big\}$,  a quasilattice vector $\{\vec{\textit{K}}_{\textit{n}}\}$ can be represented by a set of feature vectors $\{\vec{\textit{X}}_{\textit{n}}, \vec{\textit{Y}}_{\textit{n}}, \vec{\textit{Z}}_{\textit{n}},...\}$, and a feature vector $\vec{\textit{X}}_{\textit{n}}$ is composed of a set of feature components $\{\textit{x}_{\textrm{1}},..., \textit{x}_{\textit{n}}\}$. A simple example is depicted in Fig.~\ref{fig:Figure_3} $($b$)$.

  Now let the fast motion be quantum mechanical $($the evolution of matter state$)$, described by a density matrix $\hat{\rho}(\textit{t})$ driven by the Hamiltonian $\hat{\mathcal{H}}_{\textrm{0}}^{\mathds{M}}$, which is time dependent because the quasilattice point $\vec{\textit{K}}_{\textit{i}}$ changes with time. We know that there appears a magnetic reaction force at the first order, associated with the geometric phase~\cite{Wilczek-1989}. Then we assume that the gravitational-like reaction has the same expression. The evolution of $\rho$ is governed by~\cite{MVB1993}
  \begin{eqnarray}\label{commutator}
  \textit{i}\hbar\epsilon\dot{\hat{\rho}}(\textit{t}) = \big[\hat{\mathcal{H}}_{\textrm{0}}^{\mathds{M}}\big(\mathcal{K}(\textit{t})\big),\hat{\rho}(\textit{t})\big],~~\textrm{Tr}\hat{\rho}={\textrm{1}},
  \end{eqnarray}
  where $\epsilon$ is the adiabatic parameter and $\mathcal{K}(\textit{t})\propto\textit{t}$ implies ${\textit{q}}^{\textit{L}}(\mathcal{K})\propto\textit{t}$.
  The desired force is given by
  \begin{eqnarray}\label{force}
  \vec{\textit{F}} = -\textrm{Tr}\hat{\rho}~\nabla\hat{\mathcal{H}}_{\textrm{0}}^{\mathds{M}}.
  \end{eqnarray}
  One may write $\hat{\rho}$ as the series in powers of $\epsilon$
  \begin{eqnarray}\label{series}
  \hat{\rho} \equiv \sum_{\textit{r}={\textrm{0}}}^{\infty}\epsilon^{\textit{r}}\hat{\rho}_{\textit{r}}.
  \end{eqnarray}
  The terms $\hat{\rho}_{\textit{r}}$ are determined by the following equations:
  \begin{eqnarray}\label{following equations}
  \big[\hat{\mathcal{H}}_{\textrm{0}}^{\mathds{M}},\hat{\rho}_{\textrm{0}}\big] = {\textrm{0}},~~~~\big[\hat{\mathcal{H}}_{\textrm{0}}^{\mathds{M}},\hat{\rho}_{\textit{r}}\big] = i\hbar\dot{\hat{\rho}}_{\textit{r}-{\textrm{1}}}~~(\textit{r}>{\textrm{0}}).
  \end{eqnarray}
  We can choose $\hat{\rho}_{\textrm{0}}$ as one of the pure matter states as in Eq.~$($\ref{Hamiltionian3}$)$; we redefine the adiabatic eigenstates and the energy levels by
  \begin{eqnarray}\label{RedefineH}
  \hat{\mathcal{H}}_{{\textrm{0}}}^{\mathds{M}}\big(\mathcal{K}(\textit{t})\big)\big|\phi^{\textit{m}}\big(\mathcal{K}(\textit{t})\big)\big\rangle = \mathcal{E}^{\phi}\big(\mathcal{K}(\textit{t})\big)\big|\phi^{\textit{m}}\big(\mathcal{K}(\textit{t})\big)\big\rangle.
  \end{eqnarray}
  For example, the $\phi^{\textit{n}}$th is
  \begin{eqnarray}\label{rho0}
  \hat{\rho}_{\textrm{0}}(\textit{t})=\big|\phi^{\textit{n}}\big(\mathcal{K}(\textit{t})\big)\big\rangle\big\langle\phi^{\textit{n}}\big(\mathcal{K}(\textit{t})\big)\big|.
  \end{eqnarray}
  Equation~$($\ref{rho0}$)$ depends on time through varying the position of each time-dependent quasilattice function $\mathcal{K}(\textit{t})$.

  Now we can write the general force $($\ref{force}$)$ as
  \begin{eqnarray}\label{general force}
  \vec{\textit{F}} &=& -\textrm{Tr}\hat{\rho}_{\textrm{0}}\nabla\hat{\mathcal{H}}_{\textrm{0}}^{\mathds{M}}-\epsilon\textrm{Tr}\hat{\rho}_{\textrm{1}}\nabla\hat{\mathcal{H}}_{\textrm{0}}^{\mathds{M}}+\mathcal{O}
  ({\epsilon}^{\textrm{2}})\cr\cr &=& -\nabla\mathcal{E}^{\phi^{\textit{n}}}(\mathcal{K})+\epsilon\vec{\textit{F}}_{\textrm{1}} + \mathcal{O}({\epsilon}^{\textrm{2}}),
  \end{eqnarray}
  where
  \begin{eqnarray}\label{gravitational force}
  \vec{\textit{F}}_{\textrm{1}} \equiv -\textrm{Tr}\hat{\rho}_{\textrm{1}}\nabla\hat{\mathcal{H}}_{\textrm{0}}^{\mathds{M}}
  = -\sum_{{\phi}^{\textit{k}},{\phi}^{\textit{l}}}\langle{\phi}^{\textit{k}}|\hat{\rho}_{\textrm{1}}|{\phi}^{\textit{l}}\rangle\langle{\phi}^{\textit{l}}|\nabla\hat{\mathcal{H}}_{\textrm{0}}^{\mathds{M}}
  |{\phi}^{\textit{k}}\rangle.~~~~
  \end{eqnarray}
  The leading term $ -\nabla\mathcal{E}^{\phi^{\textit{n}}}$ $($also equal to $-\langle{\phi^{\textit{n}}}|\nabla\hat{\mathcal{H}}_{\textrm{0}}^{\mathds{M}}|{\phi^{\textit{n}}}\rangle$$)$ in Eq.~$($\ref{general force}$)$ is the Born-Oppenheimer force, and the second term is the desired first-order reaction. Note that $\hat{\rho}_{\textrm{1}}(\vec{\upsilon})$ is a function of slow velocity $\vec{\upsilon}\equiv\partial{\mathcal{K}}(\textit{t})/\partial\textit{t}$.

  Under the basis of adiabaticity, the off-diagonal elements of the corrections $\hat{\rho}_{\textit{r}}$ are determined by Eq.~$($\ref{following equations}$)$ as
  \begin{eqnarray}\label{matrix element}
  \langle{\phi}^{\textit{k}}|\hat{\rho}_{\textit{r}}|{\phi}^{\textit{l}}\rangle = \textit{i}\hbar\frac{\langle{\phi}^{\textit{k}}|\dot{\hat{\rho}}_{\textit{r}-{\textrm{1}}}|{\phi}^{\textit{l}}\rangle}{\mathcal{E}^{\phi^{\textit{k}}}-\mathcal{E}^{\phi^{\textit{l}}}}, ~~
  ({\phi^{\textit{k}}}\neq{\phi^{\textit{l}}}).
  \end{eqnarray}
  The diagonal elements are settled in the pure state condition $\hat{\rho}(\textit{t})=\hat{\rho}(\textit{t})^{\textrm{2}}=|\varphi(\textit{t})\rangle\langle\varphi(\textit{t})|$. Associated with Eq.~$($\ref{series}$)$, we obtain the first-order force as
  \begin{eqnarray}\label{first-order force}
  \vec{\textit{F}}_{\textrm{1}} &=& -\textit{i}\hbar\vec{\upsilon}\cdot\sum_{{\phi}^{\textit{k}},{\phi}^{\textit{l}}}\langle{\phi}^{\textit{k}}|\nabla{\phi}^{\textit{l}}\rangle(\delta_{{\phi}^{\textit{n}}{\phi}^{\textit{l}}}-
  \delta_{{\phi}^{\textit{n}}{\phi}^{\textit{k}}})\langle{\phi}^{\textit{l}}|\nabla{\phi}^{\textit{k}}\rangle\cr\cr
  &=& \textit{i}\hbar\vec{\upsilon}\wedge\sum_{{\phi}^{\textit{k}}}\langle\nabla{\phi}^{\textit{n}}|{\phi}^{\textit{k}}\rangle\wedge\langle{\phi}^{\textit{k}}|\nabla{\phi}^{\textit{n}}\rangle,
  \end{eqnarray}
  where we use $\langle{{\phi}^{\textit{l}}}|\nabla\hat{\mathcal{H}}_{\textrm{0}}^{\mathds{M}}|{{\phi}^{\textit{k}}}\rangle/(\mathcal{E}^{{\phi}^{\textit{k}}}-\mathcal{E}^{{\phi}^{\textit{l}}})
  =\langle{\phi}^{\textit{l}}|\nabla{\phi}^{\textit{k}}\rangle, ({{\phi}^{\textit{l}}}\neq{{\phi}^{\textit{k}}})$ and the completeness relation.
  This expression has the form that is analogues to Lorentz force $\vec{\textit{F}}_{\textrm{1}} = \vec{\upsilon}\wedge\vec{\textit{G}}(\mathcal{K})$, where the gravitational-like field can be denoted as $($$\hbar\equiv{\textrm{1}}$ as follows$)$
\begin{widetext}
 \begin{eqnarray}\label{magnetic field}
 \vec{\textit{G}}(\mathcal{K}) &=& -\textrm{Im}\big\langle\nabla\phi^{\textit{n}}(\mathcal{K})\big|\wedge\big|\nabla\phi^{\textit{n}}(\mathcal{K})\big\rangle\cr\cr
 &=&-\textrm{Im}\sum_{{{\phi}^{\textit{m}}}\neq{{\phi}^{\textit{n}}}}\frac{\big\langle{{\phi}^{\textit{n}}}(\mathcal{K})\big|\nabla_{\vec{\textit{K}}^{\textit{m}}}\hat{\mathcal{H}}_{\textrm{0}}^{\mathds{M}}
 (\mathcal{K})\big|{{\phi}^{\textit{m}}}(\mathcal{K})\big\rangle\big\langle{{\phi}^{\textit{m}}}(\mathcal{K})\big|\nabla_{\vec{\textit{K}}^{\textit{n}}}\hat{\mathcal{H}}_{\textrm{0}}^{\mathds{M}}(\mathcal{K})
 \big|{{\phi}^{\textit{n}}}(\mathcal{K})\big\rangle-({\vec{{\textit{K}}}^{\textit{n}}}\leftrightarrow{\vec{{\textit{K}}}^{\textit{m}}})}{\big({\mathcal{E}}^{\phi^{\textit{m}}}(\mathcal{K})-
 {\mathcal{E}}^{\phi^{\textit{n}}}(\mathcal{K})\big)^{\textrm{2}}},
 \end{eqnarray}
\end{widetext}
 which is similar to the symplectic ${\textrm{2}}$-form resulting in the geometric phase in the MG system when $\mathcal{K}$ is cycled for the gravitational field.

\begin{figure}[h]
    \centering
    \includegraphics[width=3.4in]{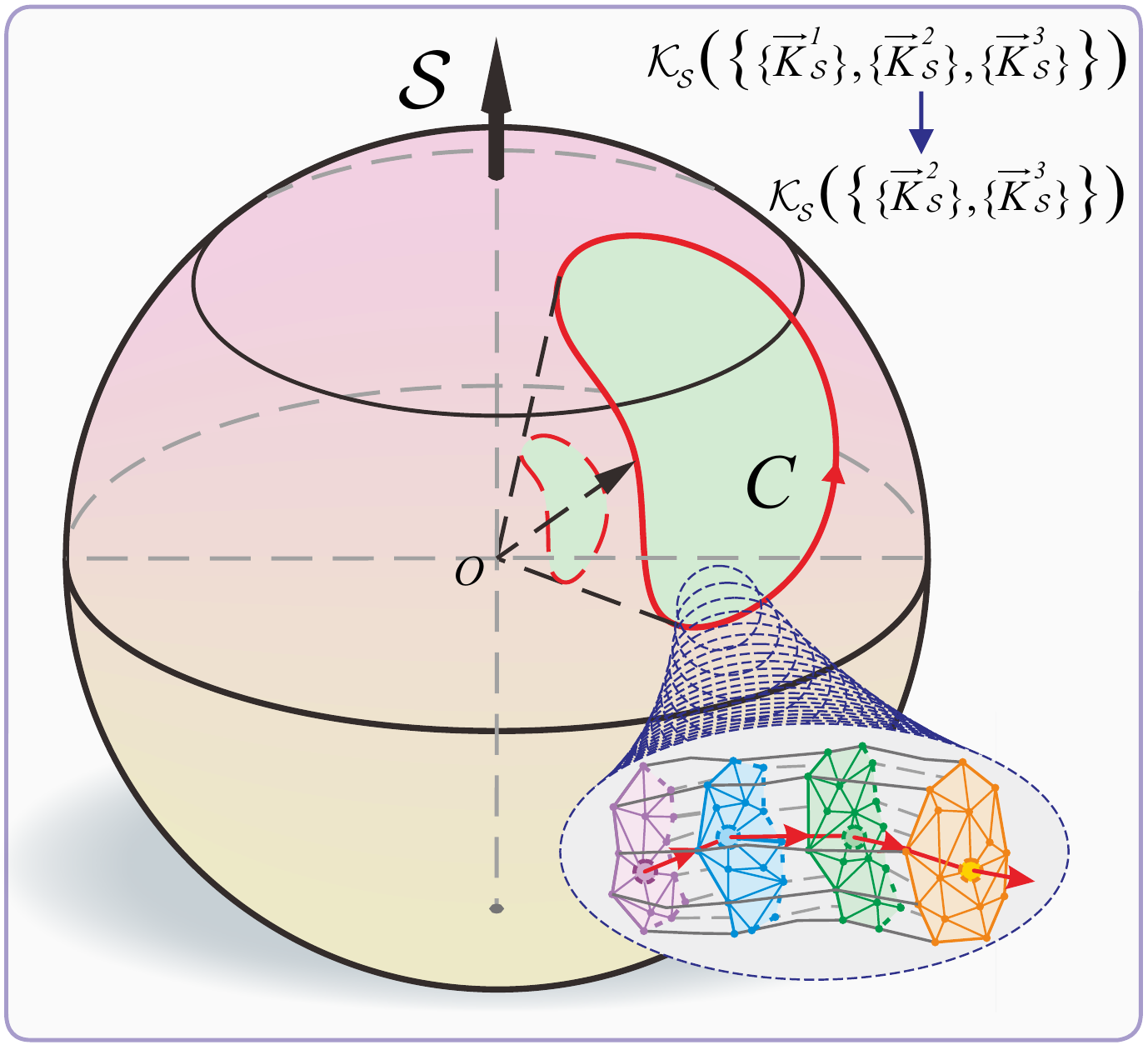}
    \caption{$($Color online$)$ A sphere constructed by the quasilattice points of the space-time. The red closed loop $\textit{C}$ represents the evolution trajectory of the matter state. The blue dashed circle depicts that $\textit{C}$ is formed by all quasilattice points during the whole evolution time, as shown in Fig.~\ref{fig:Figure_3}~$($a$)$. During this excursion along $\textit{C}$ in the external parameter space that is spanned by $\{\vec{\textit{K}}_{\mathcal{S}}^{\textrm{2}}\}$ and $\{\vec{\textit{K}}_{\mathcal{S}}^{\textrm{3}}\}$, the adiabatic change of the matter state gains a geometric phase in addition to the conventional dynamical phase.}
    \label{fig:Figure_4}
  \end{figure}

 Then we consider the topology structure of the space-time turns from $\Sigma^{\textrm{3}}\times\mathbb{R}^{\textrm{1}}$ to $\mathcal{S}^{\textrm{3}}\times\mathbb{R}^{\textrm{1}}$, where we take the simplest minisuperspace approximation into consideration. For a sphere structure of spacial lattice points, we set $\mathcal{K}_{\mathcal{S}}$ to be the quasilattice function of an emergent set of quasilattice vectors
 \begin{eqnarray}\label{position}
 \mathcal{K}_{\textit{S}} &=& \ddagger\big\{\{\vec{\textit{K}}_{\textrm{1}}^{\mathcal{S}}\},\{\vec{\textit{K}}_{\textrm{2}}^{\mathcal{S}}\}, \{\vec{\textit{K}}_{\textrm{3}}^{\mathcal{S}}\}\big\}\ddagger\cr\cr
 &=& \ddagger\big\{\{\vec{\textit{R}}_1, \vec{\Theta}_1, \vec{\Phi}_1\},...,\{\vec{\textit{R}}_n, \vec{\Theta}_n, \vec{\Phi}_n\}\big\}\ddagger\cr\cr
 &=& \big\{\{\vec{\textit{R}}_1,..., \vec{\textit{R}}_n\}, \{\vec{\Theta}_1,..., \vec{\Theta}_n\}, \{\vec{\Phi}_1,..., \vec{\Phi}_n\}\big\}\cr\cr
 &=& \big\{\{\vec{\textit{K}}_{\mathcal{S}}^{\textrm{1}}\},\{\vec{\textit{K}}_{\mathcal{S}}^{\textrm{2}}\},\{\vec{\textit{K}}_{\mathcal{S}}^{\textrm{3}}\}\big\}.
 \end{eqnarray}
%
 To ensure the only have two parameter sets of the Hamiltonian $($$\{\vec{\textit{K}}_{\mathcal{S}}^{\textrm{2}}\}$ and $\{\vec{\textit{K}}_{\mathcal{S}}^{\textrm{3}}\}$$)$, we assume that the average module of radius of the lattice sphere away from the origin during the process of the evolution of the matter state reads
 \begin{eqnarray}\label{average}
 \big\|\big\langle\{\vec{\textit{K}}_{\mathcal{S}}^{\textrm{1}}\}\big\rangle\big\|_{\textrm{0}}^{\textit{T}}=\textrm{const},~~~\textrm{or}~~~ \frac{\partial}{\partial\textit{t}}\big\langle\big\|\{\vec{\textit{K}}_{\mathcal{S}}^{\textrm{1}}\}\big\|\big\rangle\big|_{\textrm{0}}^{\textit{T}}=\textrm{0}.~~~
 \end{eqnarray}
 After this handle, the closed evolution trajectory of the matter wave function on a two-dimensional sphere in minisuperspace can be vividly pictured by Fig.~\ref{fig:Figure_4}.
 The structure of the first-order force exemplifies linear response theory, with the force $\epsilon\vec{\textit{F}}_{\textrm{1}}$ and the slow velocity $\vec{\upsilon}$ being related by the antisymmetric tensor that is the very gravitational Berry curvature in Eq.~$($\ref{imaginary_part}$)$. Thus we may conclude that the theory of half-classical approximation that introduces the notion of quasilattice points in minisuperspace is equivalent to the case of the path integral.

\section{Ripples in Hilbert space and Gravitational-like waves in minisuperspace $-$ physical quantum simulation}
\label{sec: Ripples}

  \begin{figure}[h]
    \centering
    \includegraphics[width=3.4in]{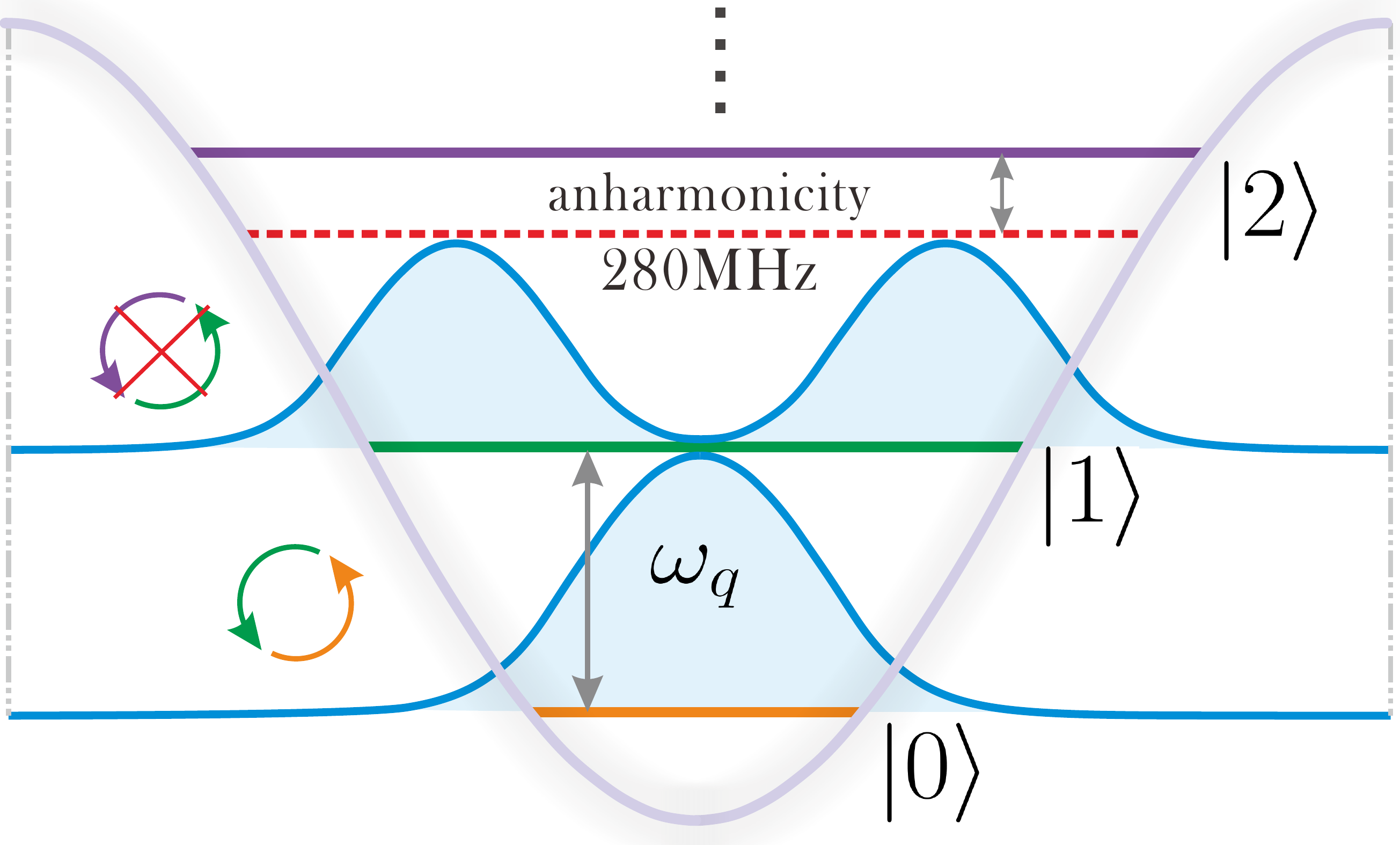}
    \caption{$($Color online$)$ Energy spectrum of the transmon qubit. Here we assume the qubit that with a transition frequency of $\omega_{\textit{q}} = \textrm{4.395}$ GHz is effectively a nonlinear resonator, and the anharmonicity of \textrm{280} MHz ensures that the qubit transition only occurs between the two lowest energy levels~\cite{KYGHSMBDGS-2007}.}
    \label{fig:Figure_5}
  \end{figure}

 A recent work studies with quantum simulation the magnetic monopoles by a driven superconducting qubit~\cite{Zhang-2016}. It shows that when the monopole charges pass from inside to outside the Hamiltonian manifold, the quantum states influenced by the Berry curvatures generate ripples in the Hilbert space.

 In order to make a comparison between the Berry curvature in the Hilbert space and the curvature of the space-time $($the gravitational Berry curvature as we mentioned before$)$ in the minisuperspace,
 we should consider a driven superconducting transmon qubit with an anharmonicity of \textrm{280} MHz forming an effective two-level system, as seen in Fig.~\ref{fig:Figure_5}. The Hamiltonian of the system can be written as
 \begin{eqnarray}\label{HHamiltonian}
 \hat{H}= \frac{\textrm{1}}{\textrm{2}}\left(
                   \begin{array}{cc}
                     \Delta & \Omega{\textit{e}}^{-\textit{i}\phi} \\
                     \Omega{\textit{e}}^{\textit{i}\phi} & -\Delta\\
                   \end{array}
                 \right),
 \end{eqnarray}
 where the detunings $\Delta = \omega_{\textit{m}} - \omega_{\textit{q}} = \Delta_{\textrm{1}}\cos\theta+\Delta_{\textrm{2}}$ $($$\omega_{\textit{m}}$ is the microwave driven frequency and $\omega_{\textit{q}}$ is the qubit transition frequency$)$, and the Rabi frequency $\Omega = \Omega_{\textrm{1}}\sin\theta$. $\phi$ is the phase of the drive tone. By changing $\Delta_{\textrm{1}}, \Delta_{\textrm{2}}$, and $\Omega_{\textrm{1}}$, we can acquire the desired single-qubit Hamiltonian that can be described in an ellipsoidal manifold spanned by these parameters. Here, we set ellipsoids of size $\Delta_{\textrm{1}} = \textrm{6}\pi \times \textrm{10}$ MHz and $\Omega = \textrm{3}\pi \times \textrm{10}$ MHz, and manipulate $\Delta_{\textrm{2}}$ from $\textrm{0}$ to $\textrm{2}\Delta_{\textrm{1}}$. It is worth noting that if $\Delta=\Omega=\textrm{0}$, this corresponds to a degeneracy in parameter space that can be viewed as a magnetic monopole~\cite{Zhang-2016}. It also can be viewed as a geometric magnetic force being that of a monopole of strength centered at $\Delta=\Omega=\textrm{0}$.

 Before we establish a connection between the monopole in parameter space and the singularity in minisuperspace, we need to show how to measure the Berry curvature. We let $\mu=\theta$ and $\nu=\phi$ in Eq.~$($\ref{Berry curvature1}$)$. So the Berry curvature is given by
 \begin{eqnarray}\label{Berry curvature}
 \textit{F}_{\theta\phi} = -\textrm{Im}\frac{\langle{\textrm{0}}|\partial_{\theta}\hat{H}|{\textrm{1}}\rangle\langle{\textrm{1}}|\partial_{\phi}\hat{H}|{\textrm{0}}\rangle -\langle {\textrm{0}}|\partial_{\phi}\hat{H}|{\textrm{1}}\rangle\langle {\textrm{1}}|\partial_{\theta}\hat{H}|{\textrm{0}}\rangle }{(\textit{E}_{\textrm{1}}-\textit{E}_{\textrm{0}})^{\textrm{2}}},\cr\cr&
 \end{eqnarray}
 where $\textit{E}_{\textit{n}}$ and $|\textit{n}\rangle$ are the $\textit{n}$th eigenvalue $($$\textit{n} = \textrm{0}, \textrm{1}$$)$ and their corresponding eigenstates of the Hamiltonian $\hat{H}$, respectively. It denotes that the local Berry curvature can be extracted from the linear response of the qubit during a nonadiabatic passage in Ref.~\cite{VGAP-2012}. This remarkable idea stems from the motion of the quantum states in a curved space deviating from a straight trajectory in flat space. Thus the Berry curvature can be calculated from the deflection from adiabaticity~\cite{PR-2014}. If we manipulate this qubit by controlling a set of parameters $($$\Delta_{\textrm{1}}, \Delta_{\textrm{2}}, \Omega_{\textrm{1}}, \theta, \phi$$)$ of its Hamiltonian with the rate of change of a parameter, then the state of the system feels a geometric $($magnetic, gravitational-like, etc.$)$ force $\textit{F}_\phi\equiv-\langle\psi(\textit{t})|\partial_{\phi}\hat{H}|\psi(\textit{t})\rangle$, given by
 \begin{equation}\label{general_force1}
 \textit{F}_\phi = \textrm{const} + \theta_{\textit{t}}\textit{F}_{\theta\phi}+\mathcal{O}({\theta}_{\textit{t}}^{\textrm{2}}),
 \end{equation}
 where $\theta_{\textit{t}}$ is the rate of change of the parameter $\theta$ $($quench velocity$)$ and $\textit{F}_{\theta\phi}$ is a component of the Berry curvature tensor. This is analogues to the case under the condition when parameters of vectors are $\vec{\textit{K}}^{\textit{m}} = \vec{\textit{K}}_{\mathcal{S}}^{\textrm{2}}$ and $\vec{\textit{K}}^{\textit{n}} = \vec{\textit{K}}_{\mathcal{S}}^{\textrm{3}}$ in Eq.~$($\ref{magnetic field}$)$. To neglect the higher-order nonlinear term $\mathcal{O}({\theta}_{\textit{t}}^{\textrm{2}})$, the system parameters should be ramped slowly enough, more details are shown in Appendix~\ref{sec: Proof2}. Note that Eq.~$($\ref{general_force1}$)$ is a specific form of Eq.~$($\ref{general force}$)$ in the case of magnetism originated from the monopole in parameter space. Analogously, gravity generated by singularity in minisuperspace can make the space-time $($spatial lattice or spatial quasilattice points$)$ curve, and the curving degree of the minisuperspace also can be described by the gravitational Berry curvature in Eq.~$($\ref{magnetic field}$)$.

 Now we reconsider the general magnetic force produced by the monopole centered at the origin of coordinates spanned by a set of parameters in the Hamiltonian. Associating with all three-space and the spatial metric tensor in space-time, we compare the singularity in minisuperspace with the magnetic monopole in parameter space. It is well known that Einstein's theory of general relativity predicts the existence of gravitational waves that are ripples in space-time at large scales, created in certain gravitational interactions that travel outward from their sources. Then we establish the relationship between the ripples characterized by the fidelity of the quantum superposition state in Hilbert space and gravitational-like waves in microscopic scales $($the mass scale of the matter is much less than the Planck mass$)$.

\begin{figure}[h]
    \centering
    \includegraphics[width=3.4in]{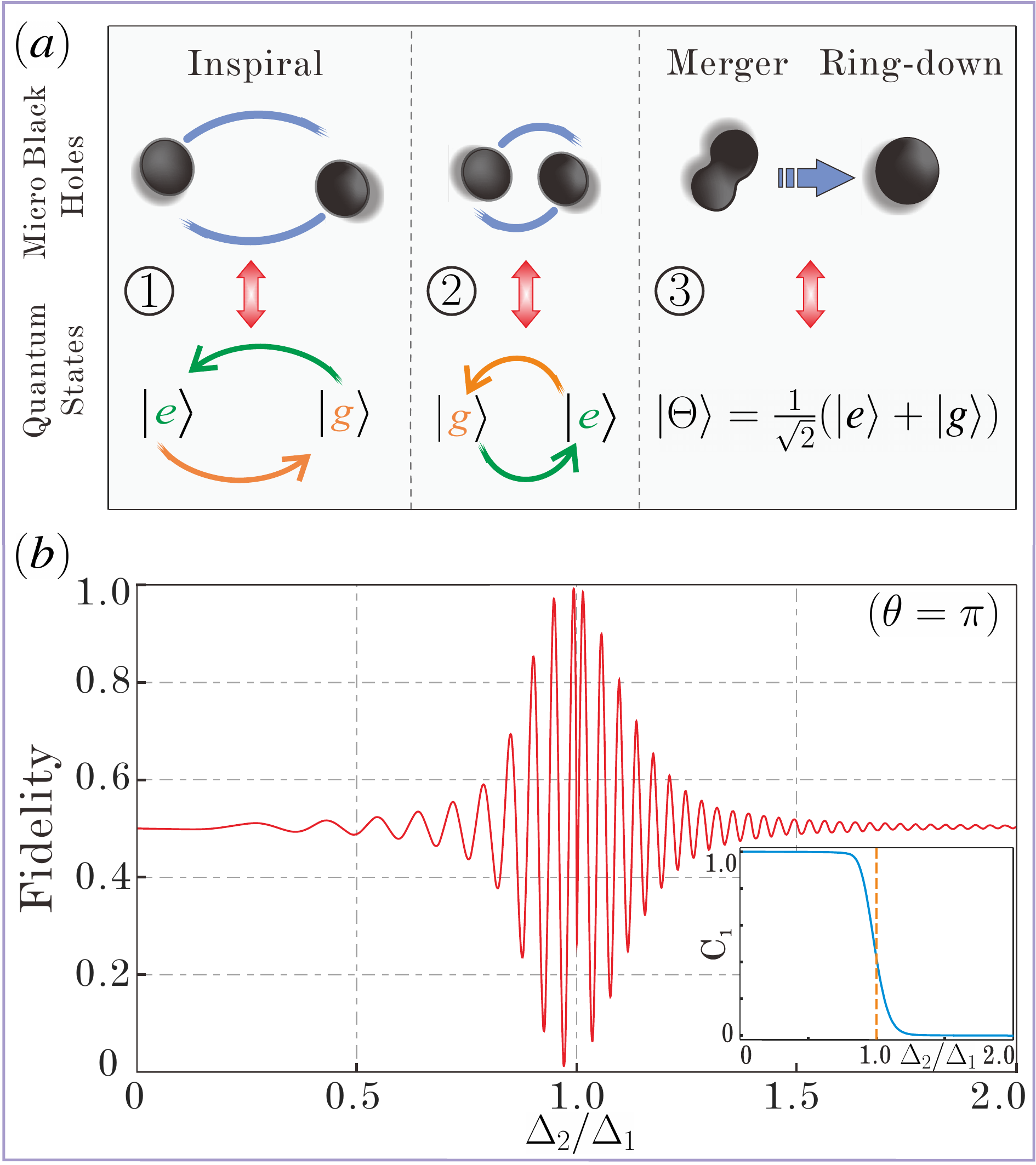}
    \caption{$($Color online$)$ Quantum simulation of artificial gravitational-like waves in minisuperspace with a two-level system. $($a$)$ The gravitational-like waves in microscopic scales emitted by the inspiral and merger of two micro black holes correspond to the process of population inversion of the two eigenstates and the process of the transfer into the superposition state, respectively. The frequency of the inspiral of micro black holes is analogues to the frequency of the population inversion of the two eigenstates. $($b$)$ Ripples of the wave function of the superposition state in Hilbert space. Here we set $\Delta_{\textrm{1}} = \textrm{6}\pi \times \textrm{10}$ MHz and $\Omega = \textrm{3}\pi \times \textrm{10}$ MHz, and manipulate $\Delta_{\textrm{2}}$ from $\textrm{0}$ to $\textrm{2}\Delta_{\textrm{1}}$ at $\theta=\pi$. The fidelity oscillates faster and faster with the increasing of $\Delta_{\textrm{2}}/\Delta_{\textrm{1}}$ before the topological transition~\cite{Zhang-2016} $($the blue real line represents the first Chern number $\textit{C}_1 = \frac{1}{2\pi}\int_0^{\pi}\textit{d}\theta\int_0^{2\pi}\textit{d}\phi\textit{F}_{\theta\phi} = \int_0^{\pi}\textit{F}_{\theta\phi}\textit{d}\theta$, which turns from 1 to 0 at $\Delta_{\textrm{2}}/\Delta_{\textrm{1}}$ in the inserted figure$)$. This corresponds to the period of inspiral of the two micro black holes becoming shorter and shorter.}
    \label{fig:Figure_6}
  \end{figure}

 We notice from Fig.~\ref{fig:Figure_6} that the number of singularity changes. In Fig.~\ref{fig:Figure_6}$($\textit{a}$)$, every micro black hole corresponds to a singularity; after the merger of two micro black holes, the number of singularities of the binary system jumps from 2 to 1. In Fig.~\ref{fig:Figure_6}$($\textit{b}$)$, after the monopole traversing outside the manifold of the parameter space, the number of singularities of the two-level system jumps from 1 to 0. These processes reflect the change of topological structures of both systems. The amplitude of gravitational-like waves from the merger of two micro black holes reaches a maximum. This instant corresponds to the moment that ripples are generated in Hilbert space when the monopole travels through $($$\Delta_{\textrm{2}}/\Delta_{\textrm{1}}=\textrm{1}$$)$ the energy surface of the Hamiltonian manifold.  In Fig.~\ref{fig:Figure_6}$($\textit{b}$)$, the fidelity of the target superposition state $|\psi\rangle={\textrm{1}}/{\sqrt{\textrm{2}}}(|\textit{g}\rangle+|\textit{e}\rangle)$ is plotted versus $\Delta_{\textrm{2}}/\Delta_{\textrm{1}}$ at $\theta=\pi$, where the fidelity of the superposition state is defined as $\textit{f} = \langle\psi|\hat{\rho}(\textit{T})|\psi\rangle$. We also note that the fidelity of the superposition state slightly oscillates around $\textrm{0.5}$ when at the region of $|\Delta_{\textrm{2}}/\Delta_{\textrm{1}}|<\textrm{0.5}$ and $|\Delta_{\textrm{2}}/\Delta_{\textrm{1}}|>\textrm{1.5}$. This implies that the quantum states almost stay in the eigenstates $($$|\textit{e}\rangle$ or $|\textit{g}\rangle$$)$ at these two regions.

%

 As shown in Fig.~\ref{fig:Figure_6}, we note that from $|\Delta_{\textrm{2}}/\Delta_{\textrm{1}}|=\textrm{0}$ to $|\Delta_{\textrm{2}}/\Delta_{\textrm{1}}|=\textrm{1}$, the ripples in Hilbert space affected by the parameter space $($energy space$)$ reflect the change of quantum states from the eigenstates $($two matter states$)$ to the superposition state. The waveform in the region of $|\Delta_{\textrm{2}}/\Delta_{\textrm{1}}|=\textrm{1}$ reflects the monopole traveling through $($$\Delta_{\textrm{2}}/\Delta_{\textrm{1}}=\textrm{1}$$)$ the energy surface of the Hamiltonian manifold; this could be viewed as the creation of gravitational-like fields by the merger of two micro black holes. When $|\Delta_{\textrm{2}}/\Delta_{\textrm{1}}|>\textrm{1}$, this afterwind could be viewed as affected by the aftershock in parameter space.

 Notice in Ref.~\cite{Zhang-2016} that, during the process of topological transition, the fidelity of the quantum states is affected by the Berry curvature, which can be expressed by
 \begin{eqnarray}\label{Measure_Berry_curvature}
 \textit{F}_{\theta\phi} = \frac{\langle\partial_\phi\hat{H}\rangle}{\theta_{\textit{t}}} = \frac{\Omega_{\textit{n}}\sin\theta}{\textrm{2}\theta_{\textit{t}}}\langle\psi|\hat{\sigma}_{\textit{y}}|\psi\rangle,
 \end{eqnarray}
 where $|\psi\rangle$ is the superposition state. It is well known that the strength of the gravitational field can be represented by the curvature of the structure of space-time. The fidelity of the quantum eigenstates is affected by the strength of the magnetic fields that are emitted from the monopoles. Then the gravitational-like interaction between the matter state and gravitational state could be described by the gravitational Berry curvature $($gravitational-like field$)$, which reflects the changes of the curvature of space-time in minisuperspace.

\section{Conclusion}
\label{sec: Conclusion}

 By use of the path integral and the half-classical approximation methods under the adiabatic approximation, we investigate the geometrical and topological structure, which are described by the gravitational Berry curvature under the theory of quantum gravity. We show that the theory of half-classical approximation that introduces the notion of quasilattice points in minisuperspace is equivalent to the case of the path integral.

 We have shown that an artificial magnetic monopole formed in parameter space of the Hamiltonian of a driven two-level system gives rise to ripples in Hilbert space when it travels through the surface of the energy manifold spanned by system parameters. The magnetic field can be represented by Berry curvature, and the distribution of Berry curvature can reflect both the shape of the manifold in parameter space of the system's Hamiltonian and the population of quantum states. Therefore, the motion of the quantum states in the curved parameter space can be utilized to stimulate the evolution of matter state in the curved minisuperspace. The first-order reaction force, also called general force, can be viewed as a characterization of the degree of Berry curvature. For example, the general magnetism force originated from the monopole in parameter space is analogues to the general gravitational-like force originated from singularity in minisuperspace; the corresponding Berry curvature $($gravitational Berry curvature$)$ reflects the strength of the magnetic field produced by the monopole and the strength of the gravitational-like field produced by the singularity $($which can be viewed as a micro black hole$)$.

 On one hand, we have devised a method of how to generate an artificial gravitational-like wave space-time for a driven superconducting qubit in the real laboratory. This might provide a feasible approach to explore the gravitational interactions between different celestial bodies through quantum simulation of the change of the curvature in space-time. On the other hand, the different number of singularities of physical systems indicates the different topological structures~\cite{MN-1998}. Thus the method of measuring the topological change in both systems may open a window for the study of the geometrical and topological properties in quantum gravity with the help of the specific quantum physical systems.

\begin{acknowledgments}

 We thank L. T. Shen for the insightful discussion. This work was supported by the National Natural Science Foundation of China under Grants  No.11405031, No.11374054, No.11305037, and No.11347114, the Major State Basic Research Development Program of China under Grant No.2012CB921601, the Natural Science Foundation of Fujian Province under Grants No.2014J05005, and the fund from Fuzhou University.

\end{acknowledgments}

\begin{appendix}

\section{Proof of Eq.~$($\ref{4Dlineelement}$)$}
\label{sec: Proof1}
  First, we introduce a spacelike hypersurface into the manifold of space-time $X^{\mu}=X^{\mu}(\textit{t},{\textit{x}}^i), \mu=0,1,2,3$ and give the normal vector ${\textit{n}^{\mu}}$ and the tangent vector $X_i^{\mu}\equiv{\partial X^{\mu}}/{\partial{\textit{x}}^i}$
  to any point on it. The local framework of four-dimensional $($${\textit{n}^{\mu}},X_i^{\mu}$$)$ satisfies the following three conditions:
\begin{description}
  \item[(1)] orthogonality: ${\textit{g}}_{\mu\nu}X_i^{\mu}{\textit{n}^{\mu}}=0$;
  \item[(2)] metric on spacelike hypersurface: ${\textit{h}}_{ij}={\textit{g}}_{\mu\nu}X_i^{\mu}X_j^{\nu}$;
  \item[(3)] timelike: ${\textit{g}}_{\mu\nu}{\textit{n}}^{\mu}{\textit{n}^{\nu}}=-1$.
\end{description}

\begin{figure}[h]
    \centering
    \includegraphics[width=2.6in]{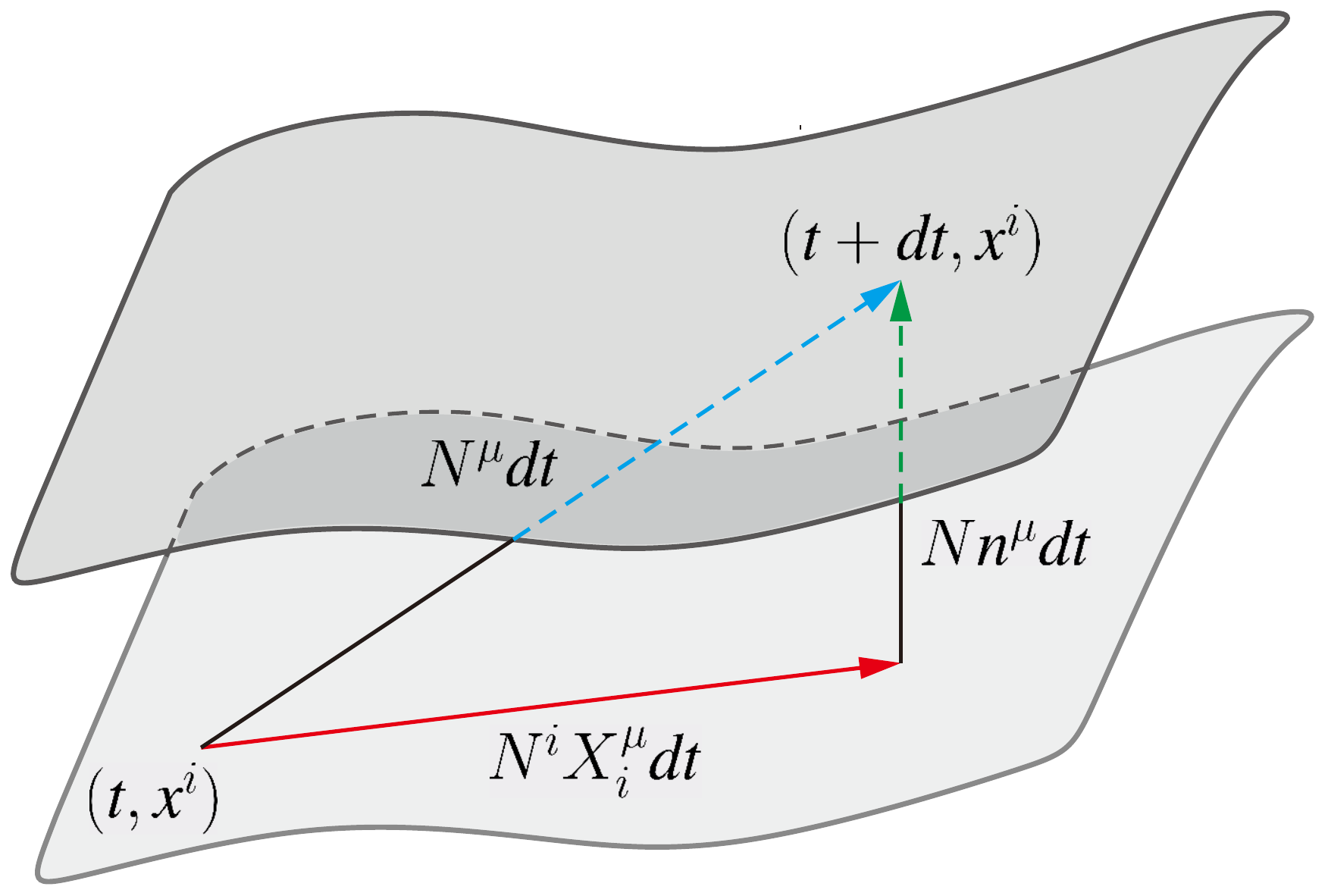}
    \caption{$($Color online$)$ Illustration of the lapse function  and the shift vector. These two quantities depict the lapse of proper time $($$N$$)$ between two infinitesimally close hypersurfaces, and the corresponding shift in spatial coordinate $($$N^i$$)$.}
    \label{fig:Figure_7}
  \end{figure}

  For spacelike hypersurface, we let $\textit{i},\textit{j}=1,2,3$. As shown in Fig.~\ref{fig:Figure_7}, continuous deformation of the hypersurface in space-time could be assumed. The deformation vector $N^{\mu}$ can be defined as $N^{\mu}\equiv{\partial X^{\mu}(\textit{t},{\textit{x}}^i)}/{\partial_t}$. In terms of a normal to the surface $\textit{n}^{\mu}$, one has
  \begin{eqnarray}\label{4frame}
  N^{\mu} = N{\textit{n}^{\mu}} + N^iX_i^{\mu}.
  \end{eqnarray}

 In Fig.~\ref{fig:Figure_7}, the space-time interval of the deformation of the hypersurface is given by
\begin{widetext}
  \begin{eqnarray}\label{4Dlineelement2}
  {\textit{ds}}^2 &=& {\textit{g}}_{\mu\nu}{\textit{d}}X^{\mu}{\textit{d}}X^{\nu} = {\textit{g}}_{\mu\nu}N^{\mu}N^{\nu}\textit{dt}\textit{dt} + 2{\textit{g}}_{\mu\nu}N^{\mu}X_i^{\nu}\textit{dt}\textit{dx}^{i} + {\textit{g}}_{\mu\nu}X_i^{\mu}X_j^{\nu}\textit{dx}^{i}\textit{dx}^{j}\cr\cr
   &=& \textit{g}_{tt}\textit{dtdt} + 2\textit{g}_{it}\textit{dx}^i\textit{dt} + \textit{g}_{ij}\textit{dx}^i\textit{dx}^j,
  \end{eqnarray}
 where ${\textit{g}_{ij}}~(i,j = 1,2,3)$ denotes the three-metric on the given spacelike hypersurface. Therefore, we have
 \begin{eqnarray}\label{4Dlineelement3}
  \textit{g}_{ij} &=& X_i^{\mu}X_j^{\nu}{\textit{g}}_{\mu\nu} = X_i^{\mu}X_j^{\nu}({\textit{h}}_{\mu\nu}-\textit{n}_{\mu}\textit{n}_{\nu}) =  X_i^{\mu}X_j^{\nu}{\textit{h}}_{\mu\nu} = \textit{h}_{ij},\cr\cr
  \textit{g}_{\textit{t}i} &=& N^{\mu}X_i^{\nu}{\textit{g}}_{\mu\nu} = {\textit{g}}_{\mu\nu}(N{\textit{n}^{\mu}} + N^jX_j^{\mu})X_i^{\nu} = N^j{\textit{g}}_{\mu\nu}X_j^{\mu}X_i^{\nu} = N^j\textit{h}_{ij} = N_i,\cr\cr
  \textit{g}_{\textit{tt}} &=& N^{\mu}N^{\nu}{\textit{g}}_{\mu\nu} = N^{\mu}N^{\nu}({\textit{h}}_{\mu\nu}-\textit{n}_{\mu}\textit{n}_{\nu}) = (N{\textit{n}^{\mu}} + N^iX_i^{\mu})(N{\textit{n}^{\nu}} + N^jX_j^{\nu}){\textit{h}}_{\mu\nu}-N^2\cr\cr
  &=& N^iX_i^{\mu}N^jX_j^{\nu}{\textit{h}}_{\mu\nu}-N^2 = N^iN_i-N^2.
 \end{eqnarray}
\end{widetext}
 This is the proof of Eq.~$($\ref{4Dlineelement}$)$.

\section{Proof of Eq.~$($\ref{general_force1}$)$}
\label{sec: Proof2}

  Now we consider a system under the domination of the Hamiltonian $\hat{H}(\textit{t}) = \hat{H}_{\textrm{0}}+{\upsilon}(\textit{t})\hat{U}$, where $\hat{H}_{\textrm{0}}$ is stationary and ${\upsilon}(\textit{t})\hat{U}$ is time dependent~\cite{CDGAP2010}. In order to characterize the dynamics of the system resulting from the time-dependent perturbation theory, we assume that ${\upsilon}(\textit{t})$ is a linear function of time,
  \begin{eqnarray}\label{linear_function}
  {\upsilon}(\textit{t}) = {\upsilon}_{\textit{i}}+\textit{t}{{\upsilon}_{\textit{t}}}({\upsilon}_{\textit{f}}-{\upsilon}_{\textit{i}}),~~~~~~\textrm{0}\leq \textit{t}\leq{\textrm{1}/{{\upsilon}_{\textit{t}}}}.
  \end{eqnarray}
  Here ${\upsilon}_{\textit{t}}$ is the rate of change of the parameter ${\upsilon}(\textit{t})$ and ${\upsilon}_{\textit{t}}\rightarrow\textrm{0}$ denotes the adiabatic limit, while
  ${\upsilon}_{\textit{t}}\rightarrow\infty$ denotes the sudden quench. The values of the vectors ${\upsilon}_{\textit{i}}$ and ${\upsilon}_{\textit{f}}$ can be arbitrarily far
  from each other in principle. Let us denote the instantaneous eigenstates of the Hamiltonian $\hat{H}(\textit{t})$ as $|\textit{n}\rangle$ satisfies the equation $\hat{H}(\textit{t})|\textit{n}\rangle = \textit{E}_{\textit{n}}(\textit{t})|\textit{n}\rangle$, where $\textit{E}_{\textit{n}}(\textit{t})$ is the corresponding instantaneous eigenvalue. We express the wave function in terms of the instantaneous eigenstates, $|\psi(\textit{t})\rangle = \sum_{\textit{n}} \textit{a}_{\textit{n}}(\textit{t})\textit{e}^{-\textit{i}\theta_{\textit{n}}(\textit{t})}|\textit{n}\rangle$, where $\theta_{\textit{n}}(\textit{t})=\int_{{\textit{t}}_{\textit{i}}}^{\textit{t}}\textit{E}_{\textit{n}}(\tau)\textit{d}\tau$ $($$\hbar\equiv\textrm{1}$$)$, and the lower limit of integration in the expression for $\theta_{\textit{n}}(\textit{t})$ is arbitrary. Then, the Schr\"{o}dinger equation reads
  \begin{equation}\label{Schrodinger1}
  \partial_{\textit{t}}{a}_{\textit{n}}(\textit{t})= -\underset{m}{\sum}\textit{a}_{\textit{m}}(\textit{t})\langle \textit{n}|\partial_{\textit{t}}|\textit{m}\rangle \textit{e}^{\textit{i}(\theta_{\textit{n}}(\textit{t})-\theta_{\textit{m}}(\textit{t}))},
  \end{equation}
  which can also be rewritten as an integral equation
  \begin{equation}\label{Schrodinger2}
  \textit{a}_{\textit{n}}(\textit{t})= -\int_{{\textit{t}}_{\textit{i}}}^{\textit{t}} {\textit{dt}}^{\prime}~\underset{\textit{m}}{\sum}\textit{a}_{\textit{m}}({\textit{t}}^{\prime})\langle \textit{n}|\partial_{{\textit{t}}^{\prime}}|\textit{m}\rangle \textit{e}^{\textit{i}(\theta_{\textit{n}}({\textit{t}}^{\prime})-\theta_{\textit{m}}({\textit{t}}^{\prime}))}.
  \end{equation}
  If the energy levels $\textit{E}_{\textit{n}}(\tau)$ and $\textit{E}_{\textit{m}}(\tau)$ are not degenerate, the matrix element $\langle \textit{n}|\partial_{\textit{t}}|\textit{m}\rangle$ can be written as
  \begin{equation}\label{adiabatic}
  \langle \textit{n}|\partial_{\textit{t}}|\textit{m}\rangle = -\frac{\langle \textit{n}|\partial_{\textit{t}}\hat{H}|\textit{m}\rangle}{\textit{E}_{\textit{n}}(\textit{t})-\textit{E}_{\textit{m}}(\textit{t})}= -\frac{{{\upsilon}_{\textit{t}}}\langle \textit{n}|\hat{U}|\textit{m}\rangle}{\textit{E}_{\textit{n}}(\textit{t})-\textit{E}_{\textit{m}}(\textit{t})}.
  \end{equation}
  If ${\upsilon}(\textit{t})$ is a monotonic function of time $\textit{t}$  then in Eq.~$($\ref{Schrodinger2}$)$ one can manipulate variables from $\textit{t}$ to ${\upsilon}(\textit{t})$ and obtain
  \begin{equation}\label{Schrodinger3}
  \textit{a}_{\textit{n}}({\upsilon})= -\int_{{\upsilon}_{\textit{i}}}^{{\upsilon}} \textit{d}{{\upsilon}}^{\prime}~\underset{\textit{m}}{\sum}\textit{a}_{\textit{m}}({{\upsilon}}^{\prime})\langle \textit{n}|\partial_{{{\upsilon}}^{\prime}}|\textit{m}\rangle \textit{e}^{\textit{i}(\theta_{\textit{n}}({{\upsilon}}^{\prime})-\theta_{\textit{m}}({{\upsilon}}^{\prime}))},
  \end{equation}
  where
  \begin{equation}\label{phase_para}
  \theta_{\textit{n}}({\upsilon})= \int_{{\upsilon}_{\textit{i}}}^{{\upsilon}}\textit{d}{{\upsilon}}^\prime \frac {\textit{E}_{\textit{n}}({\upsilon}^\prime)}{{{\upsilon}}_{\textit{t}}^\prime}.
  \end{equation}
  Equations~$($\ref{Schrodinger2}$)$ and~$($\ref{Schrodinger3}$)$ allow for a systematic expansion of the solution in the small parameter ${{\upsilon}_{\textit{t}}}$. Indeed, in the limit $\upsilon_{\textit{t}}\rightarrow \textrm{0}$ all the transition probabilities are suppressed because the phase factors are strongly oscillating functions of $\upsilon$.
  In the leading order in ${\upsilon}_{\textit{t}}$ only the term with $\textit{m} = \textit{n}$ should be retained in the sums in Eqs.~$($\ref{Schrodinger2}$)$ and~$($\ref{Schrodinger3}$)$.
  It results in the Berry phase
  \begin{equation}\label{geo_phase}
  \gamma_{\textit{n}}(\textit{t})= -\textit{i}\int_{{\textit{t}}_{\textit{i}}}^{\textit{t}}\textit{d}{\textit{t}}^\prime \langle \textit{n}|\partial_{{\textit{t}}^{\prime}}|\textit{n}\rangle = -\textit{i}\int_{{\upsilon}_{\textit{i}}}^{{\upsilon}({\textit{t}})}\textit{d}{{\upsilon}}^\prime \langle \textit{n}
  |\partial_{{\upsilon}^{\prime}}|\textit{n}\rangle.
  \end{equation}
  In general this phase could be brought into our formalism by doing a $\textit{U}$-$(\textrm{1})$ gauge transformation $\textit{a}_{\textit{n}}({\textit{t}})\rightarrow \textit{a}_{\textit{n}}({\textit{t}})\exp(-\textit{i}\gamma_{\textit{n}}({\textit{t}}))$ and changing $\theta_{\textit{n}}\rightarrow\theta_{\textit{n}}+\gamma_{\textit{n}}({\textit{t}})$ in $($\ref{Schrodinger2}$)$ and $($\ref{Schrodinger3}$)$.

  For a slow quench, ${{\upsilon}_{\textit{t}}}\ll\textrm{1}$. Given that our system is initially prepared in the ground state $\textit{n}=\textrm{0}$, so that it gives $\textit{a}_{\textrm{0}}(\textrm{0})=\textrm{1}$ and $\alpha_{\textit{n}}(\textrm{0})=\textrm{0}$ for $\textit{n}\geq\textrm{1}$. In the leading order of ${{\upsilon}_{\textit{t}}}$, Eq.~$($\ref{Schrodinger3}$)$ becomes
  \begin{equation}\label{Schrodinger4}
  \textit{a}_{\textit{n}}({\upsilon})\approx -\int_{{\upsilon}_{\textit{i}}}^{{\upsilon}} \textit{d}{{\upsilon}}^{\prime}\langle \textit{n}|\partial_{{{\upsilon}}^{\prime}}|\textrm{0}\rangle
  \textit{e}^{\textit{i}(\theta_{\textit{n}}({{\upsilon}}^{\prime})-\theta_{\textit{m}}({{\upsilon}}^{\prime}))}~.
  \end{equation}

  Now considering for a moment the one-dimensional oscillator $\textit{y}=\textit{e}^{\textit{i}\omega \textit{g}}$ we use the fact that $\textit{y}$ satisfies the differential equation
  \begin{equation}\label{asymptotic}
  {\textit{y}}^\prime(\textit{x})=\textit{i}\omega {\textit{g}}^\prime(\textit{x})\textit{y}(\textit{x}) = \textit{A}(\textit{x})\textit{y}(\textit{x}).
  \end{equation}
  The asymptotic expansion follows from writing $\textit{y}$ as $\textit{A}^{-\textrm{1}}{\textit{y}}^\prime$, assuming that $\textit{A}(\textit{x})\neq\textrm{0}$ in the interval of integration, and integrating by parts
\begin{widetext}
  \begin{eqnarray}\label{inte_part}
  \int_{\textit{a}}^{\textit{b}}\textit{fydx} = \int_{\textit{a}}^{\textit{b}}\textit{f}{\textit{A}}^{-\textrm{1}}{\textit{y}}^{\prime}\textit{dx}=
   \big[~{\textit{fA}}^{-\textrm{1}}\textit{y}\big]_{\textit{a}}^{\textit{b}}- \int_{\textit{a}}^{\textit{b}}(~{\textit{fA}}^{-\textrm{1}})^{\prime}\textit{ydx}=
   \frac{\textrm{1}}{\textit{i}\omega}\bigg(\frac{\textit{f}(\textit{b})}{{\textit{g}}^\prime(\textit{b})}\textit{y}(\textit{b})-\frac{\textit{f}(\textit{a})}{{\textit{g}}^\prime(\textit{a})}
   \textit{y}(\textit{a})\bigg)-\frac{\textrm{1}}{\textit{i}\omega}\int_{\textit{a}}^{\textit{b}}\bigg(\frac{\textit{f}}{{\textit{g}}^{\prime}}\bigg)^\prime \textit{ydx},
  \end{eqnarray}
  where the notation $\textit{A}^{-\textrm{1}}$ means matrix (or scalar) inverses, not function inverses. The first term in the right-hand side of Eq.~$($\ref{inte_part}$)$
  approximates the integral with an error
  \begin{equation}\label{omega2}
  -\frac{\textrm{1}}{\textit{i}\omega}\int_{\textit{a}}^{\textit{b}}\bigg(\frac{\textit{f}}{{\textit{g}}^{\prime}}\bigg)^\prime \textit{ydx} = \mathcal{O}(\omega^{-\textrm{2}}).
  \end{equation}
  Using the fact that the integral decays like $\mathcal{O}(\omega^{-\textrm{1}})$ \cite{ES1993}, which leads to the stand evaluation of a fast oscillating integral, $\int_{\textit{a}}^{\textit{b}}\textit{f}(\textit{x})\textit{e}^{\textit{i}\omega \textit{g}(\textit{x})}\textit{dx}=
  \frac{\textrm{1}}{\textit{i}\omega}\frac{\textit{f}(\textit{x})}{{\textit{g}}^\prime(\textit{x})}\textit{e}^{\textit{i}\omega \textit{g}(\textit{x})}\big|_{\textit{a}}^{\textit{b}}+\mathcal{O}(\omega^{-\textrm{2}})$, we obtain
  \begin{eqnarray}\label{final_int}
  \textit{a}_{\textit{n}} &\simeq&
   \textit{i}{{\upsilon}_{\textit{t}}}\frac{\langle \textit{n}|\partial_{{\upsilon}}|\textrm{0}\rangle}{\textit{E}_{\textit{n}}-\textit{E}_{\textrm{0}}}\textit{e}^{\textit{i}(\theta_{\textit{n}}({\upsilon})-\theta_{\textrm{0}}({\upsilon}))} \big|_{{\upsilon}_{\textit{i}}}^{{\upsilon}_{\textit{f}}}+\mathcal{O}({{\upsilon}}_{\textit{t}}^{\textrm{2}}) =
   -\textit{i}{{\upsilon}_{\textit{t}}}\frac{\langle \textit{n}|\partial_{{\upsilon}} \hat{H}|\textrm{0}\rangle}{(\textit{E}_{\textit{n}}-\textit{E}_{\textrm{0}})^{\textrm{2}}}\textit{e}^{\textit{i}(\theta_{\textit{n}}({\upsilon})-\theta_{\textrm{0}}({\upsilon}))} \big|_{{\upsilon}_{\textit{i}}}^{{\upsilon}_{\textit{f}}}+\mathcal{O}({{\upsilon}}_{\textit{t}}^{\textrm{2}})\cr\cr
   &=&
   -\textit{i}{\theta}_{\textit{t}}\frac{\langle \textit{n}|\partial_\theta \hat{H}|\textrm{0}\rangle}{(\textit{E}_{\textit{n}}-\textit{E}_{\textrm{0}})^{\textrm{2}}}\textit{e}^{\textit{i}(\theta_{\textit{n}}(\theta)-\theta_{\textrm{0}}(\theta))} \big|_{\theta_{\textit{i}}}^{\theta_{\textit{f}}} + \mathcal{O}({\theta}_{\textit{t}}^{\textrm{2}}) \approx
   -\textit{i}{\theta}_{\textit{t}}\frac{\langle \textit{n}|\partial_\theta \hat{H}|\textrm{0}\rangle}{(\textit{E}_{\textit{n}}-\textit{E}_{\textrm{0}})^{\textrm{2}}}\textit{e}^{-\textit{i}\theta_{\textit{n}\textrm{0}}(\theta)} \big|_{\theta_{\textit{i}}}^{\theta_{\textit{f}}},
  \end{eqnarray}
  \end{widetext}
  where ${\upsilon}_{\textit{t}} = {\theta}_{\textit{t}}$, and $\theta$ belongs to the parameter set ${\upsilon}$. $\theta_{\textit{n}\textrm{0}}$ is the full phase difference (including the dynamical and the Berry phase) between the $\textit{n}$th and the ground instantaneous eigenstates during the time evolution. Noting that Eqs.~$($\ref{phase_para}$)$ and~$($\ref{geo_phase}$)$ satisfy the $\textit{U}$-$(\textrm{1})$ gauge transformation, we get
  \begin{eqnarray}\label{theta}
  &&\theta_{\textit{n}\textrm{0}}({\upsilon})=\big(\theta_{\textit{n}}({\upsilon}^\prime)+\gamma_{\textit{n}}({\upsilon}^\prime)\big)-\big(\theta_{\textrm{0}}({\upsilon}^\prime)+\gamma_{\textrm{0}}
  ({\upsilon}^\prime)\big)\cr\cr &&=
  \int_{{\upsilon}_{\textit{i}}}^{{\upsilon}}\textit{d}{\upsilon}^{\prime}\big(\frac{\textit{E}_{\textit{n}}({\upsilon}^{\prime})-\textit{E}_{\textrm{0}}({\upsilon}^{\prime})}{{\upsilon}_{\textit{t}}
  ^{\prime}}-\big(\mathcal{A}_{\textit{n}}({\upsilon}^{\prime})-\mathcal{A}_{\textrm{0}}({\upsilon}^{\prime})\big)\big)\cr\cr
  &&=\int_{\theta_{\textit{i}}}^{\theta_{\textit{f}}} \textit{d}\theta^\prime \big(\frac{\textit{E}_{\textit{n}}(\theta^\prime)
  -\textit{E}_{\textrm{0}}(\theta^\prime)}{{\theta}_{\textit{t}}(\theta^\prime)}-\big(\mathcal{A}_{\textit{n}}(\theta^\prime)-\mathcal{A}_{\textrm{0}}(\theta^\prime)\big)\big),~~~~~
  \end{eqnarray}
  where $\mathcal{A}_{\textit{n}}$ the Berry-Simon connection equals $\textit{i}\langle \textit{n}|\partial_{{\upsilon}^{\prime}}|\textit{n}\rangle$.

  If the initial state has a large gap or if the protocol is designed in such a way that the initial evolution is adiabatic, Eq.~$($\ref{final_int}$)$ takes a particularly simple form
  \begin{equation}\label{perturbation_amplitude2}
  \textit{a}_{\textit{n}} \approx -\textit{i}{\theta}_{\textit{t}} \frac{\langle \textit{n}|\partial_\theta \hat{H}|\textrm{0}\rangle}{(\textit{E}_{\textit{n}}-\textit{E}_{\textrm{0}})^{\textrm{2}}}\big|_{\theta_{\textit{f}}}.
  \end{equation}
  The contribution of the initial term in Eq.~$($\ref{final_int}$)$ to the expectation value of the off-diagonal observables can be additionally suppressed by the fast oscillating phase $\theta_{\textit{n}\textrm{0}}$. Substituting Eq.~$($\ref{Berry curvature}$)$ into Eq.~$($\ref{perturbation_amplitude2}$)$, it is straightforward to derive the general force
\begin{widetext}
 \begin{eqnarray}\label{General_F}
  \textit{F}_{\phi} &=&
  -\langle\psi(\textit{t})|\partial_{\phi}\hat{H}|\psi(\textit{t})\rangle\approx-\langle \textrm{0}|\partial_{\phi}\hat{H}|\textrm{0}\rangle
  -{\theta}_{\textit{t}}\textrm{Im}\sum_{\textit{n}\neq\textrm{0}}\frac{\langle \textrm{0}|\partial_{\theta}\hat{H}|\textit{n}\rangle\langle \textit{n}|\partial_{\phi}\hat{H}|\textrm{0}\rangle
  -(\theta\leftrightarrow\phi)}{(\textit{E}_{\textit{n}}-\textit{E}_{\textrm{0}})^{\textrm{2}}}+\mathcal{O}(\theta_{\textit{t}}^{\textrm{2}})\cr\cr
  &=&
  \textrm{const}+{\theta}_{\textit{t}}{\textit{F}}_{\theta\phi}+\mathcal{O}(\theta_{\textit{t}}^{\textrm{2}}),
  \end{eqnarray}
  where the leading term, the Born-Oppenheimer force, is a constant, the second term is the desired first-order reaction, and
  $\mathcal{O}(\theta_{\textit{t}}^{\textrm{2}})$ indicates high-order terms that lead to derivations of the trajectory of the observables of the system.
\end{widetext}
\end{appendix}


\begin{thebibliography}{99}
\bibitem{bhmmm} In principle, a black hole can have any mass equal to or above the Planck mass.
\bibitem{string} This singular feature of the linear vector structure of the quasilattice points behaves like the process of ``peeling the onion $($layered structure$)$''. This example might reflects the emergent property of the space. We might find a more basic physical quantity to describe the notion of space that we have already known.
\end{thebibliography}
\end{document}